\documentclass[twocolumn,footinbib,preprintnumbers,floatfix,aps,prl,10pt]%
{revtex4-1}
\usepackage{amsmath,amssymb,graphicx,booktabs,bm,psfrag,color,slashed,euscript,mathtools}

\newcommand{\nn}{\nonumber}

\newcommand{\cD}{{\mathcal D}}
\newcommand{\cO}{{\mathcal O}}

\newcommand{\cT}{{\mathcal T}}

\newcommand{\cF}{{\mathcal F}}
\newcommand{\gcusp}{\gamma_{\mathrm{cusp}}}

\newcommand{\hb}{{\bar h}}
\newcommand{\fb}{{\bar f}}
\newcommand{\bmT}{{\bm T}}
\newcommand{\bmY}{{\bm Y}}

\begin{document}

\title{Infrared singularities of multi-leg QCD amplitudes with a massive parton at three loops}

\preprint{}

\author{Ze Long Liu$^a$}\email{zelong.liu@cern.ch}
\author{Nicolas Schalch$^a$}\email{schalchn@itp.unibe.ch}

\affiliation{${}^a$Institut f\"ur Theoretische Physik {\em \&} AEC, Universit\"at Bern, Sidlerstrasse 5, CH-3012 Bern, Switzerland}

\begin{abstract}
We derive the structure of three-loop anomalous dimensions governing infrared singularities of QCD amplitudes with one massive and an arbitrary number of massless external partons.  
The contributions of tripole and quadrupole correlations involving a massive parton are studied in details. The analytical expression of tripole correlations between one massive and two massless partons is obtained at three loops for the first time. We regularize the infrared divergences in the soft matrix element in a novel approach, where no extra scale dependence is involved, and the calculation can be performed in momentum space. Our results are essential to improve the theoretical predictions of single top and top quark pair productions at hadron colliders.
\end{abstract}

\maketitle

 {\em Introduction:\/} The structure of infrared (IR) singularities of gauge-theory amplitudes is of basic importance for both theory and phenomenology. 
Impressive progress has been made to investigate IR structure of multi-leg scattering amplitudes involving both massless and massive partons~\cite{Catani:1998bh,Kidonakis:1998nf,Sterman:2002qn,Bonciani:2003nt,Dokshitzer:2005ig,Aybat:2006mz,Dixon:2008gr,Becher:2009cu,Gardi:2009qi,Becher:2009qa,Dixon:2009gx,Dixon:2009ur,Becher:2009kw,Ferroglia:2009ep,Ferroglia:2009ii,Mitov:2009sv,Mitov:2010xw,DelDuca:2011ae,Caron-Huot:2013fea,Ahrens:2012qz,Almelid:2015jia,Almelid:2017qju} in the past decades. This enables us to systematically resum large logarithmic corrections to many important observables. In the LHC era, precision top quark physics is crucial for the measurements of the Standard Model parameters and for the determination of backgrounds for new physics phenomena. However,  much less is known about the IR structure of multi-leg amplitudes with massive partons beyond two-loop order, which is essential to understand soft-gluon effects and to improve the theoretical predictions for top quark productions at hadron colliders~\cite{Czakon:2009zw,Kidonakis:2010ux,Kidonakis:2011wy,Cacciari:2011hy,Ahrens:2010zv,Ahrens:2011mw,Ahrens:2011px,Ahrens:2011uf,Broggio:2016lfj,Pecjak:2016nee,Czakon:2018nun,Ju:2020otc}.
 
In soft-collinear effective theory (SCET)~\cite{Bauer:2001yt,Bauer:2002nz,Beneke:2002ph}, the IR singularities of on-shell QCD amplitudes are in one-to-one correspondence to the ultraviolet (UV) poles of low-energy matrix elements. The poles can be subtracted in $\overline{\rm MS}$ scheme by means of a multiplicative renormalization factor ${\bm Z}^{-1} (\epsilon,\{\underline{p}\},\{\underline{m}\},\mu)$. Here $\{\underline{p}\}\equiv \{p_1,p_2,...,p_n\}$ and $\{\underline{m}\}\equiv \{m_1,m_2,...,m_n\}$ denote the momenta and masses of the on-shell $n$ partons, respectively. According to the renormalization group (RG) equation,  the $Z$-factor for hard scattering amplitudes can be determined by the corresponding anomalous dimensions~\cite{Becher:2009cu,Becher:2009qa}
\begin{equation}\label{eq:Zexp}
{\bm Z} (\epsilon,\{\underline{p}\},\{\underline{m}\},\mu)= {\bf P} \exp \left[\int_\mu^\infty \frac{d\mu'}{\mu'} {\bf \Gamma}(\{\underline{p}\},\{\underline{m}\},\mu')\right],
\end{equation}
where ${\bm Z}$ and ${\bf \Gamma}$ are matrices in color space. The structure of the anomalous-dimension matrix ${\bf \Gamma}$ is severely constrained by soft-collinear factorization, non-abelian exponentiation, and the behavior of amplitudes in two-parton collinear and small-mass limits. In this Letter, we focus on the IR singularities of QCD scattering amplitudes with an arbitrary number of massless and one massive external partons. We will investigate the kinematic dependence of the tripole and quadrupole correlations in anomalous dimensions, and derive their behavior in two-particle collinear and small-mass limits. Finally, the analytical calculation for the three-loop tripole correlation involving a massive parton will be presented.

{\em General form of anomalous dimensions:\/} In SCET, the soft and collinear fields do not interact with each other after decoupling transformation. RG invariance implies that renormalization-scale dependence cancels in the combination of a hard Wilson coefficient and associated soft and collinear matrix elements. It follows that $\bm \Gamma = \bm \Gamma_s + \sum_i\Gamma_c^{i}\, \bm 1$, where $\bm \Gamma_s$ and $\Gamma_c^{i}$ denote the soft and collinear anomalous dimensions, respectively. The collinear piece can be expressed by a sum over single-parton contributions, each of which is a color-singlet and linearly depends on the collinear logarithm $L_i=\ln[\mu^2/(-p_i^2-i0)]$ \footnote{Off-shellness $p_i^2$ is applied for the $i$th massless parton to regularize IR divergences in low-energy matrix elements.} through $\Gamma_c^i=-\Gamma_{\rm cusp}^i L_i+\gamma_c^i$~\cite{Becher:2003kh}.
Here $\Gamma_{\rm cusp}^i$ is the cusp anomalous dimension depending on the color representation of parton $i$,
and $\gamma_c^i$ controls the single-logarithmic evolution.
$\bm \Gamma_s$ is a matrix in color space due to multi-parton correlations of soft interaction. The kinematic dependence of $\bm \Gamma_s$ is encoded via cusp angles formed by the Wilson lines belonging to
different pairs of massless or massive partons
\begin{gather}
\beta_{ij} = L_i+L_j-\ln\frac{\mu^2}{-s_{ij}}\,,\quad 
\beta_{Ij} = L_j-\ln\frac{m_I \mu}{-s_{Ij}}\,,\nn\\
\beta_{IJ} = {\rm cosh}^{-1}\left(\frac{-s_{IJ}}{2m_Im_J}\right)\,,
\end{gather}
where $s_{ij}=2\sigma_{ij}\ p_i\cdot p_j+i0$. The sign factor $\sigma_{ij}=+1$ if the momenta $p_i$ and $p_j$ are both incoming and outgoing, and $\sigma_{ij}=-1$ otherwise. Here and below, we label the massive partons by capital indices $I$, $J$ $\cdots$, and the massless ones by lower-cases indices $i$, $j$ $\cdots$.  Because hard scattering amplitudes are independent on the collinear scales, $L_i$ must cancel in the sum of the soft and collinear anomalous dimensions, i.e. 
\begin{align}\label{eq:scfac}
\frac{\partial {\bm \Gamma}_s}{\partial L_i} = -\frac{\partial \Gamma_c^i}{\partial L_i}\, {\bm 1}\,.
\end{align}
This implies that $\bm \Gamma_s$ can only linearly depend on the cusp angles, or depend on the conformal cross ratios of cusp angles where all the collinear scales cancel. For hard scattering amplitudes with four or more  massless external legs, the possible conformal cross ratios are found to be $\beta_{ijkl}=\beta_{ij}+\beta_{kl}-\beta_{ik}-\beta_{jl}$~\cite{Gardi:2009qi,Becher:2009qa}. According to (\ref{eq:scfac}), the coefficients of cusp angles $\beta_{ij}$ and $\beta_{Ij}$ in ${\bm \Gamma}_s$ have to be related to the cusp anomalous dimension, so the cancellation of collinear logarithms could be achieved by applying color conservation relation $\sum_i {\bm T}_i+\sum_I {\bm T}_I=0$. Here and below ${\bm T}_{i(I)}$ denotes the color generator of the $i(I)$-th massless(massive) particle in the color-space formalism of~\cite{Catani:1996jh,Catani:1996vz}.

The RG equation implies that all the UV poles of a soft matrix element can be written as an exponential of the soft anomalous dimension, in analogy to~(\ref{eq:Zexp}). Non-abelian exponentiation theorem indicates that only the maximally non-abelian part of the conventional color factor of each Feynman diagram contributes to the soft anomalous dimension. In another words, the color structures involved in soft anomalous dimensions must be connected. This was first found in the case with two Wilson lines in~\cite{Gatheral:1983cz,Frenkel:1984pz}, and then generalized to multi-parton scattering in~\cite{Gardi:2010rn,Gardi:2013ita}. 
By symmetrizing the attachments to the Wilson lines and applying the Lie algebra relation $[{\bm T}_i^a, {\bm T}_i^b]=if^{abc}{\bm T}_i^c$ repeatedly, any color structure can be rewritten as a sum of symmetric products of generators multiplied by structure constants. Eventually, soft anomalous dimensions only contains the following color structures up to three-loop order (sums over repeated color indices are implied)
\begin{gather}\label{eq:color}
\cD_{ij}={\bm T}_i^a {\bm T}_j^a\equiv {\bm T}_i\cdot {\bm T}_j \,, \quad 
\cT_{ijk}=i f^{abc} \left({\bm T}_{i}^{a} {\bm T}_{j}^{b} {\bm T}_{k}^{c}\right)_+ \,,\nn\\
\cT_{ijkl}= f^{ade} f^{bce} \left({\bm T}_{i}^{a} {\bm T}_{j}^{b} {\bm T}_{k}^{c} {\bm T}_{l}^{d}\right)_+ \,,
\end{gather} 
where $ \left({\bm T}_{i_1}^{a_1} \dots {\bm T}_{i_n}^{a_n}\right)_+ \equiv 1/n!  \sum_{\sigma} {\bm T}_{i_{\sigma(1)}}^{a_{\sigma(1)}} \dots {\bm T}_{i_{\sigma(n)}}^{a_{\sigma(n)}} $, and $\sigma$ goes through all the permutations of $n$ objects.

The structure of soft anomalous dimensions for multi-leg massless QCD amplitudes has been studied up to four-loop order. On the other hand, it is only known up to two-loop order for massive amplitudes. 
For massless amplitudes, only dipole structures $\cD_{ij}$ are involved in soft anomalous dimensions up to two-loop order, because it is impossible to construct an anti-symmetric (in the parton indices) kinematic function independent of collinear scales for the tripole structure $\cT_{ijk}$~\footnote{Throughout this letter, the tripole correlation refers to full color connections of three partons, including both $\cT_{ijk}$ and $\cT_{iijk}$ up to three-loop order. This is different from the color tripole mentioned in \cite{Almelid:2015jia,Almelid:2017qju}}. The authors in \cite{Almelid:2015jia} first calculated the non-vanishing non-dipole corrections at three-loop order, which correspond to $\cT_{ijkl}$ and are strongly constrained by two-particle collinear limits. 
For amplitudes with massive partons, $\cT_{ijk}$ associated with anti-symmetric kinematic structures can appear from two-loop order, only if at least two of the three partons are massive~\cite{Becher:2009kw,Ferroglia:2009ep}. 

In the following we will extend the structures of anomalous dimensions to multi-leg QCD amplitudes with single massive parton up to three-loop order. Color generators corresponding to massless and massive partons are mixing in color conservation, increasing the complexity of color algebra. Starting from three loops, color structures $\cT_{iiII}$, $\cT_{ijII}$, $\cT_{iijI}$ and $\cT_{ijkI}$ have to be taken into account. Color conservation implies the following identity
\begin{align}\label{eq:cTijIIrel}
{\cal T}_{ijII} = & \frac{1}{2}\left({\cal T}_{jjiI} + {\cal T}_{iijI}\right) 
- \frac{1}{2}\sum_{k\neq i,j}\left({\cal T}_{ijkI}+{\cal T}_{jikI}\right) \nn \\
&- \frac{1}{2}\sum_{J\neq I}\left({\cal T}_{ijIJ}+{\cal T}_{jiIJ}\right)
\,,
\end{align}
which helps to eliminate linearly dependent color structures in the anomalous dimensions. 
The second (third) term on the right-hand side vanishes when there are fewer than three massless (two massive) partons.
The only conformal cross ratio for kinematic functions of tripole correlations involving a massive parton is given by 
\begin{align}\label{eq:conformalr}
r_{ijI}\equiv \frac{v_I^2 \, (n_i\cdot n_j)}{2\, (v_I\cdot n_i)(v_I\cdot n_j)}\quad \mbox{with}\quad i\ne j  \,,
\end{align}
where $v_I=p_I/m_I$ is the four-velocity of massive parton $I$, and $n_{i(j)}$ is the light-like unit vector along the momentum of massless parton $i(j)$. The kinematic variables corresponding to quadrupole correlations  $\cT_{ijkI}$ can be expressed in terms of the three linearly independent variables $r_{ijI}$, $r_{ikI}$ and $r_{jkI}$, since all the other conformal ratios are fully related to these three variables, e.g. 
\begin{align}
 \frac{(n_i\cdot n_j) (v_I\cdot n_k) }{(n_i\cdot n_k) (v_I\cdot n_j)} = \frac{r_{ijI}}{r_{ikI}}\,.
\end{align}

Finally, the general structure of the three-loop anomalous dimensions for QCD amplitudes with one massive and an arbitrary number of massless partons is given by
\begin{widetext}
\begin{equation}\label{eq:gammafinalform}
\begin{aligned}
{\bf {\Gamma}}\left(\{\underline{p}\},\{\underline{m}\},\mu\right) =&
 \sum_{(i,j)}\frac{\bm{T}_i\cdot \bm{T}_j}{2}\,\gamma_{\rm cusp}(\alpha_s)\ln{\frac{\mu^2}{-s_{ij}}} 
 +\sum_{I,j} {\bm{T}_I \cdot \bm{T}_j} \,\gamma_{\rm cusp}(\alpha_s)\ln{\frac{m_I\mu}{-s_{Ij}}} 
 - \sum_{(I,J)}\frac{\bm{T}_I\cdot \bm{T}_J}{2}\,\gamma_{\rm cusp}(\beta_{IJ},\alpha_s) \\
& +\sum_{i}\gamma^i(\alpha_s)\,\bm{1} +\sum_{I}\gamma^I(\alpha_s)\,\bm{1}
 + f(\alpha_s) \sum_{(i,j,k)}\,{\cal T}_{iijk} 
 +\sum_{(i,j,k,l)}\, {\cal T}_{ijkl}\, F_{4}(\beta_{ijkl}, \beta_{ijkl}-2\beta_{ilkj},\alpha_s) \\
&  +\sum_{I}\sum_{(i,j)}\, {\cal T}_{ijII}\, F_{{\rm h}2}(r_{ijI}, \alpha_s)
 +\sum_{I}\sum_{(i,j,k)}\, {\cal T}_{ijkI}\, F_{{\rm h}3}(r_{ijI},r_{ikI},r_{jkI},\alpha_s) \\
& +[\mbox{non-dipole contributions involving two or more massive partons} ]
+\cO(\alpha_s^4)\,.
\end{aligned}
\end{equation}
\end{widetext}
Here $\gamma_{\rm cusp}(\alpha_s)$ denotes the light-like cusp anomalous dimension~\footnote{Simple Casimir scaling relation implies $\Gamma_{\rm cusp}^{i}(\alpha_s) = C_{R_i}\gamma_{\rm cusp}(\alpha_s)$, which is violated starting at four-loop order~\cite{Boels:2017skl,Boels:2017ftb}. Here $C_{R_i}={\bm T}_i^2$ is the quadratic Casimir operator of parton $i$.}, which is available up to four-loop order~\cite{Moch:2004pa,Henn:2016men,Davies:2016jie,Henn:2016wlm,Lee:2017mip,Moch:2017uml,Grozin:2018vdn,Moch:2018wjh,Lee:2019zop,Henn:2019rmi,vonManteuffel:2019wbj,Henn:2019swt,vonManteuffel:2020vjv,Agarwal:2021zft}.  $\gamma_{\rm cusp}(\beta_{IJ},\alpha_s)$ is the angle-dependent cusp anomalous dimension, which has been fully obtained up to three-loop order in QCD~\cite{Kidonakis:2009ev,Grozin:2014hna,Grozin:2015kna}, and is partially known at four loops~\cite{Bruser:2019auj,Bruser:2020bsh}.
The collinear anomalous dimensions $\gamma^{q(g)}$ can be extracted from the divergent part of the quark (gluon) form factor up to four loops~\cite{Moch:2005id,Moch:2005tm,Baikov:2009bg,vonManteuffel:2020vjv,Agarwal:2021zft}.  $\gamma^Q$ is available up to three loops~\cite{Korchemsky:1987wg,Korchemsky:1991zp,Kidonakis:2009ev,Grozin:2014hna,Grozin:2015kna,Bruser:2019yjk}. 
The contributions in the first two lines of eq.~(\ref{eq:gammafinalform}) have been presented in~\cite{Becher:2009cu,Gardi:2009qi,Becher:2009qa,Almelid:2015jia,Becher:2019avh}. The terms in the third line denote tripole and quadrupole correlations with a massive parton starting from three-loop order. Due to symmetry properties of $\cT_{ijkl}$ and $\beta_{ijkl}$~\cite{Becher:2009qa}, $F_4$ and $F_{\rm h3}$ can be chosen as odd functions, i.e. $F_4(x,y,\alpha_s)=-F_4(-x,y,\alpha_s)$ and $F_{\rm h3}(x,y,z,\alpha_s)=-F_{\rm h3}(y,x,z,\alpha_s)$. 
The kinematic functions in~(\ref{eq:gammafinalform}) are strongly constrained by the small-mass limits. When the masses of the external partons are much smaller than the characteristic hard scales, the amplitude can factorize into a product of jet functions, describing collinear singularities, times the corresponding massless amplitude~\cite{Mitov:2006xs,Becher:2007cu}.  
This implies that there is no color exchange between different external partons in ${\bf {\Gamma}}\left(\{\underline{p}\},\{\underline{m}\to 0\},\mu\right)-{\bf {\Gamma}}\left(\{\underline{p}\},\{\underline{0}\},\mu\right)$, where ${\bf {\Gamma}}\left(\{\underline{p}\},\{\underline{0}\},\mu\right)$ denotes the corresponding anomalous dimension in purely massless case. 
Moreover, the anomalous dimensions for $1\to 2$ splitting amplitudes ${\bf \Gamma}_{\rm Sp}$ can be determined by ${\bf {\Gamma}}\left(\{\underline{p}\},\{\underline{m}\},\mu\right)$ in~(\ref{eq:gammafinalform}) when the momenta of any two massless particles are aligned. The fact that ${\bf \Gamma}_{\rm Sp}$ only depends on color generators for the two daughter particles requires that the contributions involving color generators for other particles must cancel out. 
The relevant derivations are provided in the supplemental material. As a result, we have following relations
\begin{gather}\label{eq:ffuncrel1}
\lim_{\omega\to -\infty}F_4(\omega,\omega,\alpha_s)= \frac{f(\alpha_s)}{2}\,, \\
F_{\rm h2}(0,\alpha_s)=  3f(\alpha_s)\,, \qquad 
F_{\rm h3}(0,r,r,\alpha_s) =  2 f(\alpha_s)\,,\nn
\end{gather}
and
\begin{align}\label{eq:ffuncrel2}
&\lim_{v_I^2\to 0}F_{\rm h3}(r_{ijI},r_{ikI},r_{jkI},\alpha_s) \nn\\
& =  2 f(\alpha_s)+  4 F_{4}(\beta_{ijkI},\beta_{ijkI}-2\beta_{kjiI},\alpha_s)\,.
\end{align}
The first relation in~(\ref{eq:ffuncrel1}) was first obtained in~\cite{Almelid:2015jia}, while the others are derived for the first time in this letter.

{\em Calculation of $F_{\rm h2}$:\/} 
\begin{figure}
\begin{center}
\includegraphics[width=0.44\textwidth]{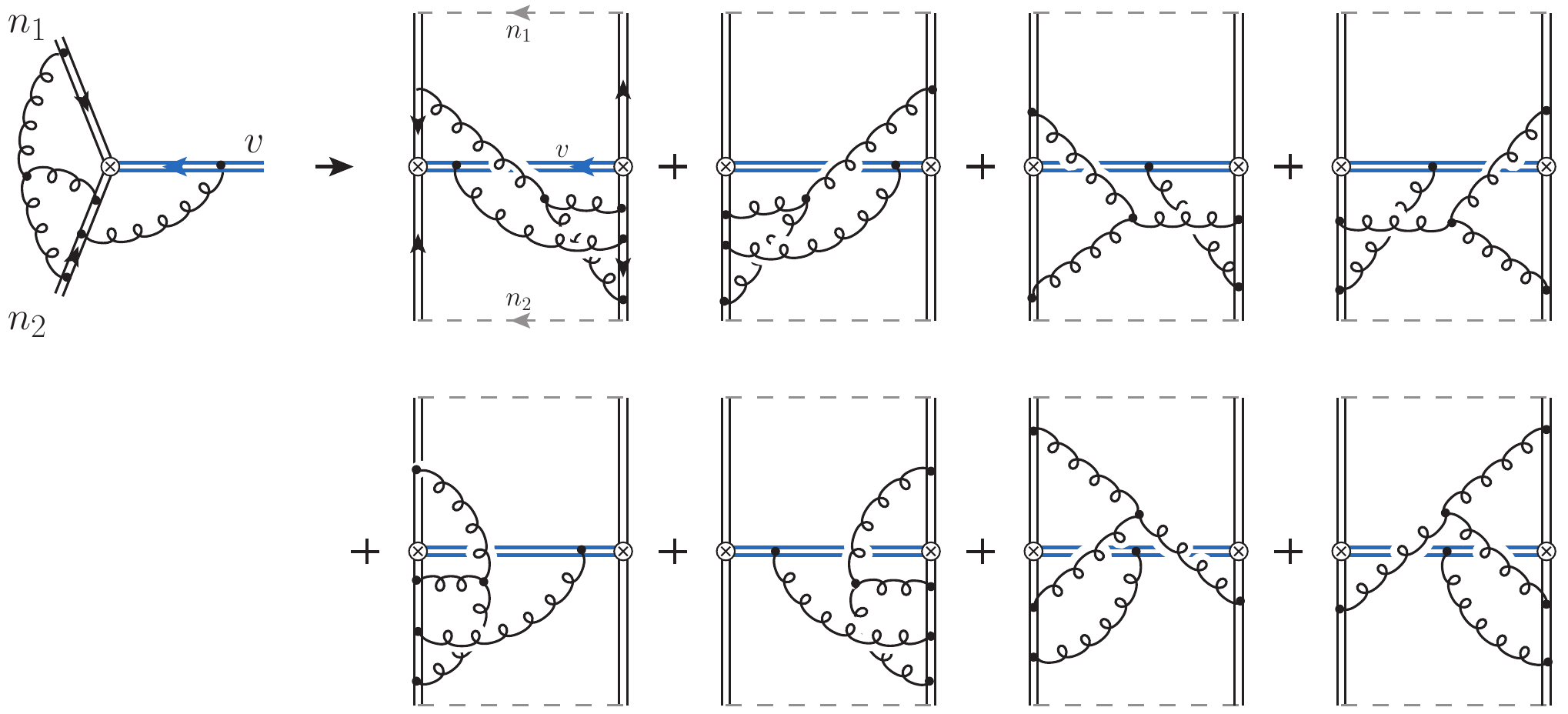}	
\caption{An example for the relation of color structures between diagrams of the soft correlator and the soft function in~(\ref{eq:softfunc2}). For the subdiagrams on right hand side, the external and internal double lines correspond to the semi-infinite and finite-length soft Wilson lines in~(\ref{eq:softfunc2a}), respectively. The dashed gray lines denote color connections between the soft Wilson lines along the same directions.}
\label{fig:fig1}
\end{center}
\end{figure}
In this section, we present the calculation of the three-loop coefficient
\begin{align}
F_{{\rm h}2}(r, \alpha_s)=& \left(\frac{\alpha_s}{4\pi}\right)^3 {\cal F}_{{\rm h}2}(r) + {\cal O}(\alpha_s^4)\,.
\end{align}
${\cal F}_{{\rm h}2}(r)$ can be conveniently obtained through the evaluation of soft anomalous dimensions. 
To extract the UV poles of soft matrix elements, appropriate regulators need to be introduced to regularize IR divergences. For example, an exponential regulator was proposed in~\cite{Gardi:2013saa} to isolate UV poles, and the relevant calculations were performed systematically in configuration space. In~\cite{Ferroglia:2009ii}, IR divergences are regularized by assigning a residual external momentum to each Wilson line. We note that for physical observables, the low-energy matrix elements in SCET are free of IR poles because they are regularized by the low-energy measurements. This provides a natural approach to isolate UV poles of soft matrix elements. Specifically, we consider the soft function in factorization at cross-section level
\begin{equation}\label{eq:softfunc1}
{\bm S} (\omega) = \langle 0| \overline {\rm T}\left[ {\bm Y}^\dagger_{n_1} {\bm Y}^\dagger_{n_2} {\bm Y}^\dagger_{v} \right]
\delta(\omega - v\cdot \hat p)  {\rm T}\left[ {\bm Y}_{n_1} {\bm Y}_{n_2} {\bm Y}_{v} \right]|0\rangle \,,
\end{equation}   
where $\bm Y_{n(v)}=\bm Y_{n(v)}(0)$ denotes semi-infinite soft Wilson line along $n^\mu(v^\mu)$ direction, ${\hat p}^\mu$ is the momentum operator picking up the total momentum of all soft emissions in final states, and ${\rm T}(\overline{\rm T})$ indicates (anti-)time ordering. In practice, this soft function has applications in phenomenology. For instance, it can describe the soft-gluon effects for near-threshold production of single top quark associated with color-singlet states (e.g $W$ boson or charge Higgs) at hadron colliders.
In~\cite{Liu:2020wmp} one of us and a collaborator have provided a novel method to compute inclusive soft functions in terms of loop diagrams. 
In particular, the soft function defined in~(\ref{eq:softfunc1}) can be rewritten as
\begin{equation}\label{eq:softfunc2}
\begin{aligned}
{\bm S}(\omega) = & \frac{1}{2\pi}{\rm Re} \left[{\bm \Sigma}(\omega+i0)- {\bm \Sigma}(\omega-i0)\right]\,,
\end{aligned}
\end{equation}
with 
\begin{equation}\label{eq:softfunc2a}
\begin{aligned}
{\bm \Sigma}(\omega)=&\int_0^{\infty}\!\! dt\,  e^{i\omega t}
 \langle 0| {\rm T}\Big[\bmY_{n_1}^\dagger (t v)\bmY_{n_2}^\dagger (t v) \\
& \cdot {\rm P} \exp \Big[ig\int_0^t \!\! ds \, v\cdot A^c(sv)\bmT_v^c\Big]
\bmY_{n_1} (0)\bmY_{n_2} (0)\Big] |0 \rangle\,,
\end{aligned}
\end{equation} 
where $\rm P$ indicates path ordering. This allowed us to avoid any phase space integrations and to straightforwardly take advantage of well-established multi-loop technology. 

In our calculation, only color-connected diagrams, also called  {\it webs} are taken into account due to non-abelian exponentiation theorem. Unlike the soft correlator $\langle 0| \bmY_{n_1} \bmY_{n_2} \bmY_{v} |0\rangle$, the soft function in~(\ref{eq:softfunc2}) appears in factorizations at cross-section level. Nevertheless, the replica trick for evaluating the diagrammatic contributions to the exponent~\cite{Gardi:2010rn,Gardi:2013ita} is still compatible~\footnote{The web mixing matrices are even available at four loops~\cite{Agarwal:2020nyc,Agarwal:2021him}}.
A sample is shown in fig.~\ref{fig:fig1}, where each subdiagram on right hand side has the same color structure as the left one. This can be seen by moving the gluon endpoints sequentially from right to left along the attached Wilson lines in each subdiagram. There are several advantages to extract UV poles by evaluating diagrams from definition in~(\ref{eq:softfunc2a}). First, $\omega$ is the only dimensionful kinematic variable in the integrals, so it can factor out and does not increase the complexity of the integrals. Although both IR and UV poles exist in individual diagrams, all the IR poles cancel out when summing over all the diagrams contributing to $\cF_{\rm h2}(r)$. Second, gauge invariance is preserved, and the calculation can be performed in general covariant gauge. Finally, the calculation can be performed in momentum space, which allows us to use sophisticated multi-loop computation techniques, e.g. integration-by-parts (IBP) reduction and the differential equation (DE) method.

The calculation is performed in dimensional regularization $d=4-2\epsilon$ and in general covariant gauge with gauge parameter $\xi$. We use $\texttt{QGRAF}$~\cite{Nogueira:1991ex} to generate the color connected diagrams at three loops. After partial-fraction decompositions, the scalar Feynman integrals in the diagrams can be mapped onto thirty integral topologies, each of which consists of fifteen linearly independent quadratic and linear propagators.
Using IBP reduction and eliminating redundant MIs across the integral topologies~\cite{Smirnov:2019qkx,Klappert:2020nbg}, $\cF_{\rm h2}(r)$ can be further expressed as a linear combination of 173 linearly independent master integrals (MIs). In this expression the gauge parameter $\xi$ manifestly cancels out, demonstrating the validity of our setup. In the next step, we use the DE method to solve for the MIs. The public packages $\texttt{CANONICA}$~\cite{Meyer:2017joq} and $\texttt{DlogBasis}$~\cite{Henn:2020lye} are helpful to convert the DE systems into a canonical form~\cite{Henn:2013pwa}. The resulting symbol alphabet is $\{r,r-1,r-2,(r-1)\sqrt{r},\sqrt{r(r-1)}\}$, where the last letter leads to the generalized harmonic polylogarithms (GHPLs)~\cite{Aglietti:2004tq} in the solution of the DEs, and the letter $(r-1)\sqrt{r}$ can be rationalized by changing variables to $u=\sqrt{r}-1$. The boundary conditions are determined by the values of the MIs at $r=1$, which corresponds to the kinematic point $v^\mu=n_1^\mu + n_2^\mu$. The dimensional recurrence relations~\cite{Tarasov:1996br,Lee:2009dh,Lee:2013mka} help to express each MI at $r=1$ in terms of a set of quasi-finite integrals in $d=n-2\epsilon$ $(n=4,6,8,\dots)$, which can be evaluated by performing the integrations over the Feynman parameters with the package $\texttt{HyperInt}$~\cite{Panzer:2014caa}. Eventually, we can iteratively solve the DEs order-by-order in $\epsilon$ in terms of Goncharov Polylogarithms (GPLs) and GHPLs.

After inserting the results of the MIs into the expression of $\cF_{\rm h2}(r)$, all the poles from $\epsilon^{-5}$ down to $\epsilon^{-1}$ notably cancel out. Furthermore, all the GHPLs also manifestly drop out. 
Finally, the expression can be remarkably simplified to
\begin{widetext}
\begin{equation}\label{eq:finalFh2r}
\begin{aligned}
{\cal F}_{{\rm h}2}(r) =& 
128 \Big[H_{-1,0,0,0} +H_{-1,1,0,0} +H_{1,-1,0,0} -H_{1,0,0,0} \Big] 
+128 \left(\zeta _2+\zeta _3\right) \Big[H_{1,0} -H_{-1,0} \Big]
+96 \left(\zeta _3 + \zeta _4\right) \Big[H_{-1} -H_1 \Big]
\\&
+128 \zeta _2 \Big[H_{-2,0} -H_{2,0} +H_{-1,0,0} -H_{1,0,0} \Big]
+256 \Big[H_{1,2,0,0} +H_{2,0,0,0} -H_{-2,0,0,0} +H_{-1,-2,0,0} -H_{-1,2,0,0} 
\\&\qquad\quad
-H_{1,-2,0,0} -H_{-1,0,0,0,0} +H_{1,0,0,0,0}  \Big]
+48 \left(2 \zeta _2 \zeta _3+\zeta _5\right) \,,
\end{aligned}
\end{equation}
\end{widetext}
where $H_{\vec {a}}\equiv H_{\vec {a}}(\sqrt{r})$ are the harmonic polylogarithms (HPLs)~\cite{Remiddi:1999ew,Maitre:2005uu}.
We use the notation of dropping the zeros in the vector $\vec a$, adding 1 to the absolute value of the next right non-zero index for each dropped 0. In small-mass limit $r\to 0$, all the terms in~(\ref{eq:finalFh2r}) vanish except the last one, which returns to the tripole contribution in the purely massless cases, as shown in~(\ref{eq:ffuncrel1}). $F_{{\rm h}2}(r,\alpha_s)$ does not have a uniform transcendental weight $(2L-1)$ at $L$ loops, differing from the tripole and quadrupole correlations in purely massless cases. This interesting observation has also been found recently in the boomerang-type webs~\cite{Gardi:2021gzz}.

{\em Summary:\/} Based on soft-collinear factorization and non-abelian exponentiation theorem, we have derived the general form of anomalous dimensions governing IR singularities of QCD amplitudes with one massive and an arbitrary number of massless partons up to three-loop order. In comparison to the purely massless cases, two additional color structures are introduced, and the corresponding kinematic variables have been determined. We discuss the relations between the kinematic coefficients using the constraints from small-mass and two-particle collinear limits. The three-loop analytical expression of the tripole correlation involving a massive parton has been obtained for the first time, which can be directly used to improve theoretical predictions of cross sections for single top productions. It is also an important ingredient to the IR singulariteis of QCD amplitudes with a heavy quark pair.

\vspace{1mm}
{\em Acknowledgements:\/}
We are grateful to Thomas Becher for many stimulating discussions and for a careful proofreading of the manuscript, and to Xiaofeng Xu for helpful dicussions.  Z.L.L thanks Robin Br\"user for providing the codes to perform the topology mapping and partial fractioning.
The research is supported by the Swiss National Science Foundation (SNF) under grant 200020\_182038.

\bibliographystyle{apsrev4-1}
\bibliography{letter}

\begin{thebibliography}{96}%
\makeatletter
\providecommand \@ifxundefined [1]{%
 \@ifx{#1\undefined}
}%
\providecommand \@ifnum [1]{%
 \ifnum #1\expandafter \@firstoftwo
 \else \expandafter \@secondoftwo
 \fi
}%
\providecommand \@ifx [1]{%
 \ifx #1\expandafter \@firstoftwo
 \else \expandafter \@secondoftwo
 \fi
}%
\providecommand \natexlab [1]{#1}%
\providecommand \enquote  [1]{``#1''}%
\providecommand \bibnamefont  [1]{#1}%
\providecommand \bibfnamefont [1]{#1}%
\providecommand \citenamefont [1]{#1}%
\providecommand \href@noop [0]{\@secondoftwo}%
\providecommand \href [0]{\begingroup \@sanitize@url \@href}%
\providecommand \@href[1]{\@@startlink{#1}\@@href}%
\providecommand \@@href[1]{\endgroup#1\@@endlink}%
\providecommand \@sanitize@url [0]{\catcode `\\12\catcode `\$12\catcode
  `\&12\catcode `\#12\catcode `\^12\catcode `\_12\catcode `\%12\relax}%
\providecommand \@@startlink[1]{}%
\providecommand \@@endlink[0]{}%
\providecommand \url  [0]{\begingroup\@sanitize@url \@url }%
\providecommand \@url [1]{\endgroup\@href {#1}{\urlprefix }}%
\providecommand \urlprefix  [0]{URL }%
\providecommand \Eprint [0]{\href }%
\providecommand \doibase [0]{http://dx.doi.org/}%
\providecommand \selectlanguage [0]{\@gobble}%
\providecommand \bibinfo  [0]{\@secondoftwo}%
\providecommand \bibfield  [0]{\@secondoftwo}%
\providecommand \translation [1]{[#1]}%
\providecommand \BibitemOpen [0]{}%
\providecommand \bibitemStop [0]{}%
\providecommand \bibitemNoStop [0]{.\EOS\space}%
\providecommand \EOS [0]{\spacefactor3000\relax}%
\providecommand \BibitemShut  [1]{\csname bibitem#1\endcsname}%
\let\auto@bib@innerbib\@empty
\bibitem [{\citenamefont {Catani}(1998)}]{Catani:1998bh}%
  \BibitemOpen
  \bibfield  {author} {\bibinfo {author} {\bibfnamefont {S.}~\bibnamefont
  {Catani}},\ }\href {\doibase 10.1016/S0370-2693(98)00332-3} {\bibfield
  {journal} {\bibinfo  {journal} {Phys. Lett. B}\ }\textbf {\bibinfo {volume}
  {427}},\ \bibinfo {pages} {161} (\bibinfo {year} {1998})},\ \Eprint
  {http://arxiv.org/abs/hep-ph/9802439} {arXiv:hep-ph/9802439} \BibitemShut
  {NoStop}%
\bibitem [{\citenamefont {Kidonakis}\ \emph {et~al.}(1998)\citenamefont
  {Kidonakis}, \citenamefont {Oderda},\ and\ \citenamefont
  {Sterman}}]{Kidonakis:1998nf}%
  \BibitemOpen
  \bibfield  {author} {\bibinfo {author} {\bibfnamefont {N.}~\bibnamefont
  {Kidonakis}}, \bibinfo {author} {\bibfnamefont {G.}~\bibnamefont {Oderda}}, \
  and\ \bibinfo {author} {\bibfnamefont {G.~F.}\ \bibnamefont {Sterman}},\
  }\href {\doibase 10.1016/S0550-3213(98)00441-6} {\bibfield  {journal}
  {\bibinfo  {journal} {Nucl. Phys. B}\ }\textbf {\bibinfo {volume} {531}},\
  \bibinfo {pages} {365} (\bibinfo {year} {1998})},\ \Eprint
  {http://arxiv.org/abs/hep-ph/9803241} {arXiv:hep-ph/9803241} \BibitemShut
  {NoStop}%
\bibitem [{\citenamefont {Sterman}\ and\ \citenamefont
  {Tejeda-Yeomans}(2003)}]{Sterman:2002qn}%
  \BibitemOpen
  \bibfield  {author} {\bibinfo {author} {\bibfnamefont {G.~F.}\ \bibnamefont
  {Sterman}}\ and\ \bibinfo {author} {\bibfnamefont {M.~E.}\ \bibnamefont
  {Tejeda-Yeomans}},\ }\href {\doibase 10.1016/S0370-2693(02)03100-3}
  {\bibfield  {journal} {\bibinfo  {journal} {Phys. Lett. B}\ }\textbf
  {\bibinfo {volume} {552}},\ \bibinfo {pages} {48} (\bibinfo {year} {2003})},\
  \Eprint {http://arxiv.org/abs/hep-ph/0210130} {arXiv:hep-ph/0210130}
  \BibitemShut {NoStop}%
\bibitem [{\citenamefont {Bonciani}\ \emph {et~al.}(2003)\citenamefont
  {Bonciani}, \citenamefont {Catani}, \citenamefont {Mangano},\ and\
  \citenamefont {Nason}}]{Bonciani:2003nt}%
  \BibitemOpen
  \bibfield  {author} {\bibinfo {author} {\bibfnamefont {R.}~\bibnamefont
  {Bonciani}}, \bibinfo {author} {\bibfnamefont {S.}~\bibnamefont {Catani}},
  \bibinfo {author} {\bibfnamefont {M.~L.}\ \bibnamefont {Mangano}}, \ and\
  \bibinfo {author} {\bibfnamefont {P.}~\bibnamefont {Nason}},\ }\href
  {\doibase 10.1016/j.physletb.2003.09.068} {\bibfield  {journal} {\bibinfo
  {journal} {Phys. Lett. B}\ }\textbf {\bibinfo {volume} {575}},\ \bibinfo
  {pages} {268} (\bibinfo {year} {2003})},\ \Eprint
  {http://arxiv.org/abs/hep-ph/0307035} {arXiv:hep-ph/0307035} \BibitemShut
  {NoStop}%
\bibitem [{\citenamefont {Dokshitzer}\ and\ \citenamefont
  {Marchesini}(2006)}]{Dokshitzer:2005ig}%
  \BibitemOpen
  \bibfield  {author} {\bibinfo {author} {\bibfnamefont {Y.~L.}\ \bibnamefont
  {Dokshitzer}}\ and\ \bibinfo {author} {\bibfnamefont {G.}~\bibnamefont
  {Marchesini}},\ }\href {\doibase 10.1088/1126-6708/2006/01/007} {\bibfield
  {journal} {\bibinfo  {journal} {JHEP}\ }\textbf {\bibinfo {volume} {01}},\
  \bibinfo {pages} {007} (\bibinfo {year} {2006})},\ \Eprint
  {http://arxiv.org/abs/hep-ph/0509078} {arXiv:hep-ph/0509078} \BibitemShut
  {NoStop}%
\bibitem [{\citenamefont {Aybat}\ \emph {et~al.}(2006)\citenamefont {Aybat},
  \citenamefont {Dixon},\ and\ \citenamefont {Sterman}}]{Aybat:2006mz}%
  \BibitemOpen
  \bibfield  {author} {\bibinfo {author} {\bibfnamefont {S.~M.}\ \bibnamefont
  {Aybat}}, \bibinfo {author} {\bibfnamefont {L.~J.}\ \bibnamefont {Dixon}}, \
  and\ \bibinfo {author} {\bibfnamefont {G.~F.}\ \bibnamefont {Sterman}},\
  }\href {\doibase 10.1103/PhysRevD.74.074004} {\bibfield  {journal} {\bibinfo
  {journal} {Phys. Rev. D}\ }\textbf {\bibinfo {volume} {74}},\ \bibinfo
  {pages} {074004} (\bibinfo {year} {2006})},\ \Eprint
  {http://arxiv.org/abs/hep-ph/0607309} {arXiv:hep-ph/0607309} \BibitemShut
  {NoStop}%
\bibitem [{\citenamefont {Dixon}\ \emph {et~al.}(2008)\citenamefont {Dixon},
  \citenamefont {Magnea},\ and\ \citenamefont {Sterman}}]{Dixon:2008gr}%
  \BibitemOpen
  \bibfield  {author} {\bibinfo {author} {\bibfnamefont {L.~J.}\ \bibnamefont
  {Dixon}}, \bibinfo {author} {\bibfnamefont {L.}~\bibnamefont {Magnea}}, \
  and\ \bibinfo {author} {\bibfnamefont {G.~F.}\ \bibnamefont {Sterman}},\
  }\href {\doibase 10.1088/1126-6708/2008/08/022} {\bibfield  {journal}
  {\bibinfo  {journal} {JHEP}\ }\textbf {\bibinfo {volume} {08}},\ \bibinfo
  {pages} {022} (\bibinfo {year} {2008})},\ \Eprint
  {http://arxiv.org/abs/0805.3515} {arXiv:0805.3515 [hep-ph]} \BibitemShut
  {NoStop}%
\bibitem [{\citenamefont {Becher}\ and\ \citenamefont
  {Neubert}(2009{\natexlab{a}})}]{Becher:2009cu}%
  \BibitemOpen
  \bibfield  {author} {\bibinfo {author} {\bibfnamefont {T.}~\bibnamefont
  {Becher}}\ and\ \bibinfo {author} {\bibfnamefont {M.}~\bibnamefont
  {Neubert}},\ }\href {\doibase 10.1103/PhysRevLett.102.162001,
  10.1103/PhysRevLett.111.199905} {\bibfield  {journal} {\bibinfo  {journal}
  {Phys. Rev. Lett.}\ }\textbf {\bibinfo {volume} {102}},\ \bibinfo {pages}
  {162001} (\bibinfo {year} {2009}{\natexlab{a}})},\ \bibinfo {note} {[Erratum:
  Phys. Rev. Lett.111,no.19,199905(2013)]},\ \Eprint
  {http://arxiv.org/abs/0901.0722} {arXiv:0901.0722 [hep-ph]} \BibitemShut
  {NoStop}%
\bibitem [{\citenamefont {Gardi}\ and\ \citenamefont
  {Magnea}(2009)}]{Gardi:2009qi}%
  \BibitemOpen
  \bibfield  {author} {\bibinfo {author} {\bibfnamefont {E.}~\bibnamefont
  {Gardi}}\ and\ \bibinfo {author} {\bibfnamefont {L.}~\bibnamefont {Magnea}},\
  }\href {\doibase 10.1088/1126-6708/2009/03/079} {\bibfield  {journal}
  {\bibinfo  {journal} {JHEP}\ }\textbf {\bibinfo {volume} {03}},\ \bibinfo
  {pages} {079} (\bibinfo {year} {2009})},\ \Eprint
  {http://arxiv.org/abs/0901.1091} {arXiv:0901.1091 [hep-ph]} \BibitemShut
  {NoStop}%
\bibitem [{\citenamefont {Becher}\ and\ \citenamefont
  {Neubert}(2009{\natexlab{b}})}]{Becher:2009qa}%
  \BibitemOpen
  \bibfield  {author} {\bibinfo {author} {\bibfnamefont {T.}~\bibnamefont
  {Becher}}\ and\ \bibinfo {author} {\bibfnamefont {M.}~\bibnamefont
  {Neubert}},\ }\href {\doibase 10.1088/1126-6708/2009/06/081,
  10.1007/JHEP11(2013)024} {\bibfield  {journal} {\bibinfo  {journal} {JHEP}\
  }\textbf {\bibinfo {volume} {06}},\ \bibinfo {pages} {081} (\bibinfo {year}
  {2009}{\natexlab{b}})},\ \bibinfo {note} {[Erratum: JHEP11,024(2013)]},\
  \Eprint {http://arxiv.org/abs/0903.1126} {arXiv:0903.1126 [hep-ph]}
  \BibitemShut {NoStop}%
\bibitem [{\citenamefont {Dixon}(2009)}]{Dixon:2009gx}%
  \BibitemOpen
  \bibfield  {author} {\bibinfo {author} {\bibfnamefont {L.~J.}\ \bibnamefont
  {Dixon}},\ }\href {\doibase 10.1103/PhysRevD.79.091501} {\bibfield  {journal}
  {\bibinfo  {journal} {Phys. Rev. D}\ }\textbf {\bibinfo {volume} {79}},\
  \bibinfo {pages} {091501(R)} (\bibinfo {year} {2009})},\ \Eprint
  {http://arxiv.org/abs/0901.3414} {arXiv:0901.3414 [hep-ph]} \BibitemShut
  {NoStop}%
\bibitem [{\citenamefont {Dixon}\ \emph {et~al.}(2010)\citenamefont {Dixon},
  \citenamefont {Gardi},\ and\ \citenamefont {Magnea}}]{Dixon:2009ur}%
  \BibitemOpen
  \bibfield  {author} {\bibinfo {author} {\bibfnamefont {L.~J.}\ \bibnamefont
  {Dixon}}, \bibinfo {author} {\bibfnamefont {E.}~\bibnamefont {Gardi}}, \ and\
  \bibinfo {author} {\bibfnamefont {L.}~\bibnamefont {Magnea}},\ }\href
  {\doibase 10.1007/JHEP02(2010)081} {\bibfield  {journal} {\bibinfo  {journal}
  {JHEP}\ }\textbf {\bibinfo {volume} {02}},\ \bibinfo {pages} {081} (\bibinfo
  {year} {2010})},\ \Eprint {http://arxiv.org/abs/0910.3653} {arXiv:0910.3653
  [hep-ph]} \BibitemShut {NoStop}%
\bibitem [{\citenamefont {Becher}\ and\ \citenamefont
  {Neubert}(2009{\natexlab{c}})}]{Becher:2009kw}%
  \BibitemOpen
  \bibfield  {author} {\bibinfo {author} {\bibfnamefont {T.}~\bibnamefont
  {Becher}}\ and\ \bibinfo {author} {\bibfnamefont {M.}~\bibnamefont
  {Neubert}},\ }\href {\doibase 10.1103/PhysRevD.79.125004} {\bibfield
  {journal} {\bibinfo  {journal} {Phys. Rev. D}\ }\textbf {\bibinfo {volume}
  {79}},\ \bibinfo {pages} {125004} (\bibinfo {year} {2009}{\natexlab{c}})},\
  \bibinfo {note} {[Erratum: Phys.Rev.D 80, 109901(E) (2009)]},\ \Eprint
  {http://arxiv.org/abs/0904.1021} {arXiv:0904.1021 [hep-ph]} \BibitemShut
  {NoStop}%
\bibitem [{\citenamefont {Ferroglia}\ \emph
  {et~al.}(2009{\natexlab{a}})\citenamefont {Ferroglia}, \citenamefont
  {Neubert}, \citenamefont {Pecjak},\ and\ \citenamefont
  {Yang}}]{Ferroglia:2009ep}%
  \BibitemOpen
  \bibfield  {author} {\bibinfo {author} {\bibfnamefont {A.}~\bibnamefont
  {Ferroglia}}, \bibinfo {author} {\bibfnamefont {M.}~\bibnamefont {Neubert}},
  \bibinfo {author} {\bibfnamefont {B.~D.}\ \bibnamefont {Pecjak}}, \ and\
  \bibinfo {author} {\bibfnamefont {L.~L.}\ \bibnamefont {Yang}},\ }\href
  {\doibase 10.1103/PhysRevLett.103.201601} {\bibfield  {journal} {\bibinfo
  {journal} {Phys. Rev. Lett.}\ }\textbf {\bibinfo {volume} {103}},\ \bibinfo
  {pages} {201601} (\bibinfo {year} {2009}{\natexlab{a}})},\ \Eprint
  {http://arxiv.org/abs/0907.4791} {arXiv:0907.4791 [hep-ph]} \BibitemShut
  {NoStop}%
\bibitem [{\citenamefont {Ferroglia}\ \emph
  {et~al.}(2009{\natexlab{b}})\citenamefont {Ferroglia}, \citenamefont
  {Neubert}, \citenamefont {Pecjak},\ and\ \citenamefont
  {Yang}}]{Ferroglia:2009ii}%
  \BibitemOpen
  \bibfield  {author} {\bibinfo {author} {\bibfnamefont {A.}~\bibnamefont
  {Ferroglia}}, \bibinfo {author} {\bibfnamefont {M.}~\bibnamefont {Neubert}},
  \bibinfo {author} {\bibfnamefont {B.~D.}\ \bibnamefont {Pecjak}}, \ and\
  \bibinfo {author} {\bibfnamefont {L.~L.}\ \bibnamefont {Yang}},\ }\href
  {\doibase 10.1088/1126-6708/2009/11/062} {\bibfield  {journal} {\bibinfo
  {journal} {JHEP}\ }\textbf {\bibinfo {volume} {11}},\ \bibinfo {pages} {062}
  (\bibinfo {year} {2009}{\natexlab{b}})},\ \Eprint
  {http://arxiv.org/abs/0908.3676} {arXiv:0908.3676 [hep-ph]} \BibitemShut
  {NoStop}%
\bibitem [{\citenamefont {Mitov}\ \emph {et~al.}(2009)\citenamefont {Mitov},
  \citenamefont {Sterman},\ and\ \citenamefont {Sung}}]{Mitov:2009sv}%
  \BibitemOpen
  \bibfield  {author} {\bibinfo {author} {\bibfnamefont {A.}~\bibnamefont
  {Mitov}}, \bibinfo {author} {\bibfnamefont {G.~F.}\ \bibnamefont {Sterman}},
  \ and\ \bibinfo {author} {\bibfnamefont {I.}~\bibnamefont {Sung}},\ }\href
  {\doibase 10.1103/PhysRevD.79.094015} {\bibfield  {journal} {\bibinfo
  {journal} {Phys. Rev. D}\ }\textbf {\bibinfo {volume} {79}},\ \bibinfo
  {pages} {094015} (\bibinfo {year} {2009})},\ \Eprint
  {http://arxiv.org/abs/0903.3241} {arXiv:0903.3241 [hep-ph]} \BibitemShut
  {NoStop}%
\bibitem [{\citenamefont {Mitov}\ \emph {et~al.}(2010)\citenamefont {Mitov},
  \citenamefont {Sterman},\ and\ \citenamefont {Sung}}]{Mitov:2010xw}%
  \BibitemOpen
  \bibfield  {author} {\bibinfo {author} {\bibfnamefont {A.}~\bibnamefont
  {Mitov}}, \bibinfo {author} {\bibfnamefont {G.~F.}\ \bibnamefont {Sterman}},
  \ and\ \bibinfo {author} {\bibfnamefont {I.}~\bibnamefont {Sung}},\ }\href
  {\doibase 10.1103/PhysRevD.82.034020} {\bibfield  {journal} {\bibinfo
  {journal} {Phys. Rev. D}\ }\textbf {\bibinfo {volume} {82}},\ \bibinfo
  {pages} {034020} (\bibinfo {year} {2010})},\ \Eprint
  {http://arxiv.org/abs/1005.4646} {arXiv:1005.4646 [hep-ph]} \BibitemShut
  {NoStop}%
\bibitem [{\citenamefont {Del~Duca}\ \emph {et~al.}(2011)\citenamefont
  {Del~Duca}, \citenamefont {Duhr}, \citenamefont {Gardi}, \citenamefont
  {Magnea},\ and\ \citenamefont {White}}]{DelDuca:2011ae}%
  \BibitemOpen
  \bibfield  {author} {\bibinfo {author} {\bibfnamefont {V.}~\bibnamefont
  {Del~Duca}}, \bibinfo {author} {\bibfnamefont {C.}~\bibnamefont {Duhr}},
  \bibinfo {author} {\bibfnamefont {E.}~\bibnamefont {Gardi}}, \bibinfo
  {author} {\bibfnamefont {L.}~\bibnamefont {Magnea}}, \ and\ \bibinfo {author}
  {\bibfnamefont {C.~D.}\ \bibnamefont {White}},\ }\href {\doibase
  10.1007/JHEP12(2011)021} {\bibfield  {journal} {\bibinfo  {journal} {JHEP}\
  }\textbf {\bibinfo {volume} {12}},\ \bibinfo {pages} {021} (\bibinfo {year}
  {2011})},\ \Eprint {http://arxiv.org/abs/1109.3581} {arXiv:1109.3581
  [hep-ph]} \BibitemShut {NoStop}%
\bibitem [{\citenamefont {Caron-Huot}(2015)}]{Caron-Huot:2013fea}%
  \BibitemOpen
  \bibfield  {author} {\bibinfo {author} {\bibfnamefont {S.}~\bibnamefont
  {Caron-Huot}},\ }\href {\doibase 10.1007/JHEP05(2015)093} {\bibfield
  {journal} {\bibinfo  {journal} {JHEP}\ }\textbf {\bibinfo {volume} {05}},\
  \bibinfo {pages} {093} (\bibinfo {year} {2015})},\ \Eprint
  {http://arxiv.org/abs/1309.6521} {arXiv:1309.6521 [hep-th]} \BibitemShut
  {NoStop}%
\bibitem [{\citenamefont {Ahrens}\ \emph {et~al.}(2012)\citenamefont {Ahrens},
  \citenamefont {Neubert},\ and\ \citenamefont {Vernazza}}]{Ahrens:2012qz}%
  \BibitemOpen
  \bibfield  {author} {\bibinfo {author} {\bibfnamefont {V.}~\bibnamefont
  {Ahrens}}, \bibinfo {author} {\bibfnamefont {M.}~\bibnamefont {Neubert}}, \
  and\ \bibinfo {author} {\bibfnamefont {L.}~\bibnamefont {Vernazza}},\ }\href
  {\doibase 10.1007/JHEP09(2012)138} {\bibfield  {journal} {\bibinfo  {journal}
  {JHEP}\ }\textbf {\bibinfo {volume} {09}},\ \bibinfo {pages} {138} (\bibinfo
  {year} {2012})},\ \Eprint {http://arxiv.org/abs/1208.4847} {arXiv:1208.4847
  [hep-ph]} \BibitemShut {NoStop}%
\bibitem [{\citenamefont {Almelid}\ \emph {et~al.}(2016)\citenamefont
  {Almelid}, \citenamefont {Duhr},\ and\ \citenamefont
  {Gardi}}]{Almelid:2015jia}%
  \BibitemOpen
  \bibfield  {author} {\bibinfo {author} {\bibfnamefont {O.}~\bibnamefont
  {Almelid}}, \bibinfo {author} {\bibfnamefont {C.}~\bibnamefont {Duhr}}, \
  and\ \bibinfo {author} {\bibfnamefont {E.}~\bibnamefont {Gardi}},\ }\href
  {\doibase 10.1103/PhysRevLett.117.172002} {\bibfield  {journal} {\bibinfo
  {journal} {Phys. Rev. Lett.}\ }\textbf {\bibinfo {volume} {117}},\ \bibinfo
  {pages} {172002} (\bibinfo {year} {2016})},\ \Eprint
  {http://arxiv.org/abs/1507.00047} {arXiv:1507.00047 [hep-ph]} \BibitemShut
  {NoStop}%
\bibitem [{\citenamefont {Almelid}\ \emph {et~al.}(2017)\citenamefont
  {Almelid}, \citenamefont {Duhr}, \citenamefont {Gardi}, \citenamefont
  {McLeod},\ and\ \citenamefont {White}}]{Almelid:2017qju}%
  \BibitemOpen
  \bibfield  {author} {\bibinfo {author} {\bibfnamefont {O.}~\bibnamefont
  {Almelid}}, \bibinfo {author} {\bibfnamefont {C.}~\bibnamefont {Duhr}},
  \bibinfo {author} {\bibfnamefont {E.}~\bibnamefont {Gardi}}, \bibinfo
  {author} {\bibfnamefont {A.}~\bibnamefont {McLeod}}, \ and\ \bibinfo {author}
  {\bibfnamefont {C.~D.}\ \bibnamefont {White}},\ }\href {\doibase
  10.1007/JHEP09(2017)073} {\bibfield  {journal} {\bibinfo  {journal} {JHEP}\
  }\textbf {\bibinfo {volume} {09}},\ \bibinfo {pages} {073} (\bibinfo {year}
  {2017})},\ \Eprint {http://arxiv.org/abs/1706.10162} {arXiv:1706.10162
  [hep-ph]} \BibitemShut {NoStop}%
\bibitem [{\citenamefont {Czakon}\ \emph {et~al.}(2009)\citenamefont {Czakon},
  \citenamefont {Mitov},\ and\ \citenamefont {Sterman}}]{Czakon:2009zw}%
  \BibitemOpen
  \bibfield  {author} {\bibinfo {author} {\bibfnamefont {M.}~\bibnamefont
  {Czakon}}, \bibinfo {author} {\bibfnamefont {A.}~\bibnamefont {Mitov}}, \
  and\ \bibinfo {author} {\bibfnamefont {G.~F.}\ \bibnamefont {Sterman}},\
  }\href {\doibase 10.1103/PhysRevD.80.074017} {\bibfield  {journal} {\bibinfo
  {journal} {Phys. Rev. D}\ }\textbf {\bibinfo {volume} {80}},\ \bibinfo
  {pages} {074017} (\bibinfo {year} {2009})},\ \Eprint
  {http://arxiv.org/abs/0907.1790} {arXiv:0907.1790 [hep-ph]} \BibitemShut
  {NoStop}%
\bibitem [{\citenamefont {Kidonakis}(2010)}]{Kidonakis:2010ux}%
  \BibitemOpen
  \bibfield  {author} {\bibinfo {author} {\bibfnamefont {N.}~\bibnamefont
  {Kidonakis}},\ }\href {\doibase 10.1103/PhysRevD.82.054018} {\bibfield
  {journal} {\bibinfo  {journal} {Phys. Rev. D}\ }\textbf {\bibinfo {volume}
  {82}},\ \bibinfo {pages} {054018} (\bibinfo {year} {2010})},\ \Eprint
  {http://arxiv.org/abs/1005.4451} {arXiv:1005.4451 [hep-ph]} \BibitemShut
  {NoStop}%
\bibitem [{\citenamefont {Kidonakis}(2011)}]{Kidonakis:2011wy}%
  \BibitemOpen
  \bibfield  {author} {\bibinfo {author} {\bibfnamefont {N.}~\bibnamefont
  {Kidonakis}},\ }\href {\doibase 10.1103/PhysRevD.83.091503} {\bibfield
  {journal} {\bibinfo  {journal} {Phys. Rev. D}\ }\textbf {\bibinfo {volume}
  {83}},\ \bibinfo {pages} {091503(R)} (\bibinfo {year} {2011})},\ \Eprint
  {http://arxiv.org/abs/1103.2792} {arXiv:1103.2792 [hep-ph]} \BibitemShut
  {NoStop}%
\bibitem [{\citenamefont {Cacciari}\ \emph {et~al.}(2012)\citenamefont
  {Cacciari}, \citenamefont {Czakon}, \citenamefont {Mangano}, \citenamefont
  {Mitov},\ and\ \citenamefont {Nason}}]{Cacciari:2011hy}%
  \BibitemOpen
  \bibfield  {author} {\bibinfo {author} {\bibfnamefont {M.}~\bibnamefont
  {Cacciari}}, \bibinfo {author} {\bibfnamefont {M.}~\bibnamefont {Czakon}},
  \bibinfo {author} {\bibfnamefont {M.}~\bibnamefont {Mangano}}, \bibinfo
  {author} {\bibfnamefont {A.}~\bibnamefont {Mitov}}, \ and\ \bibinfo {author}
  {\bibfnamefont {P.}~\bibnamefont {Nason}},\ }\href {\doibase
  10.1016/j.physletb.2012.03.013} {\bibfield  {journal} {\bibinfo  {journal}
  {Phys. Lett. B}\ }\textbf {\bibinfo {volume} {710}},\ \bibinfo {pages} {612}
  (\bibinfo {year} {2012})},\ \Eprint {http://arxiv.org/abs/1111.5869}
  {arXiv:1111.5869 [hep-ph]} \BibitemShut {NoStop}%
\bibitem [{\citenamefont {Ahrens}\ \emph {et~al.}(2010)\citenamefont {Ahrens},
  \citenamefont {Ferroglia}, \citenamefont {Neubert}, \citenamefont {Pecjak},\
  and\ \citenamefont {Yang}}]{Ahrens:2010zv}%
  \BibitemOpen
  \bibfield  {author} {\bibinfo {author} {\bibfnamefont {V.}~\bibnamefont
  {Ahrens}}, \bibinfo {author} {\bibfnamefont {A.}~\bibnamefont {Ferroglia}},
  \bibinfo {author} {\bibfnamefont {M.}~\bibnamefont {Neubert}}, \bibinfo
  {author} {\bibfnamefont {B.~D.}\ \bibnamefont {Pecjak}}, \ and\ \bibinfo
  {author} {\bibfnamefont {L.~L.}\ \bibnamefont {Yang}},\ }\href {\doibase
  10.1007/JHEP09(2010)097} {\bibfield  {journal} {\bibinfo  {journal} {JHEP}\
  }\textbf {\bibinfo {volume} {09}},\ \bibinfo {pages} {097} (\bibinfo {year}
  {2010})},\ \Eprint {http://arxiv.org/abs/1003.5827} {arXiv:1003.5827
  [hep-ph]} \BibitemShut {NoStop}%
\bibitem [{\citenamefont {Ahrens}\ \emph
  {et~al.}(2011{\natexlab{a}})\citenamefont {Ahrens}, \citenamefont
  {Ferroglia}, \citenamefont {Neubert}, \citenamefont {Pecjak},\ and\
  \citenamefont {Yang}}]{Ahrens:2011mw}%
  \BibitemOpen
  \bibfield  {author} {\bibinfo {author} {\bibfnamefont {V.}~\bibnamefont
  {Ahrens}}, \bibinfo {author} {\bibfnamefont {A.}~\bibnamefont {Ferroglia}},
  \bibinfo {author} {\bibfnamefont {M.}~\bibnamefont {Neubert}}, \bibinfo
  {author} {\bibfnamefont {B.~D.}\ \bibnamefont {Pecjak}}, \ and\ \bibinfo
  {author} {\bibfnamefont {L.-L.}\ \bibnamefont {Yang}},\ }\href {\doibase
  10.1007/JHEP09(2011)070} {\bibfield  {journal} {\bibinfo  {journal} {JHEP}\
  }\textbf {\bibinfo {volume} {09}},\ \bibinfo {pages} {070} (\bibinfo {year}
  {2011}{\natexlab{a}})},\ \Eprint {http://arxiv.org/abs/1103.0550}
  {arXiv:1103.0550 [hep-ph]} \BibitemShut {NoStop}%
\bibitem [{\citenamefont {Ahrens}\ \emph
  {et~al.}(2011{\natexlab{b}})\citenamefont {Ahrens}, \citenamefont
  {Ferroglia}, \citenamefont {Neubert}, \citenamefont {Pecjak},\ and\
  \citenamefont {Yang}}]{Ahrens:2011px}%
  \BibitemOpen
  \bibfield  {author} {\bibinfo {author} {\bibfnamefont {V.}~\bibnamefont
  {Ahrens}}, \bibinfo {author} {\bibfnamefont {A.}~\bibnamefont {Ferroglia}},
  \bibinfo {author} {\bibfnamefont {M.}~\bibnamefont {Neubert}}, \bibinfo
  {author} {\bibfnamefont {B.~D.}\ \bibnamefont {Pecjak}}, \ and\ \bibinfo
  {author} {\bibfnamefont {L.~L.}\ \bibnamefont {Yang}},\ }\href {\doibase
  10.1016/j.physletb.2011.07.058} {\bibfield  {journal} {\bibinfo  {journal}
  {Phys. Lett. B}\ }\textbf {\bibinfo {volume} {703}},\ \bibinfo {pages} {135}
  (\bibinfo {year} {2011}{\natexlab{b}})},\ \Eprint
  {http://arxiv.org/abs/1105.5824} {arXiv:1105.5824 [hep-ph]} \BibitemShut
  {NoStop}%
\bibitem [{\citenamefont {Ahrens}\ \emph
  {et~al.}(2011{\natexlab{c}})\citenamefont {Ahrens}, \citenamefont
  {Ferroglia}, \citenamefont {Neubert}, \citenamefont {Pecjak},\ and\
  \citenamefont {Yang}}]{Ahrens:2011uf}%
  \BibitemOpen
  \bibfield  {author} {\bibinfo {author} {\bibfnamefont {V.}~\bibnamefont
  {Ahrens}}, \bibinfo {author} {\bibfnamefont {A.}~\bibnamefont {Ferroglia}},
  \bibinfo {author} {\bibfnamefont {M.}~\bibnamefont {Neubert}}, \bibinfo
  {author} {\bibfnamefont {B.~D.}\ \bibnamefont {Pecjak}}, \ and\ \bibinfo
  {author} {\bibfnamefont {L.~L.}\ \bibnamefont {Yang}},\ }\href {\doibase
  10.1103/PhysRevD.84.074004} {\bibfield  {journal} {\bibinfo  {journal} {Phys.
  Rev. D}\ }\textbf {\bibinfo {volume} {84}},\ \bibinfo {pages} {074004}
  (\bibinfo {year} {2011}{\natexlab{c}})},\ \Eprint
  {http://arxiv.org/abs/1106.6051} {arXiv:1106.6051 [hep-ph]} \BibitemShut
  {NoStop}%
\bibitem [{\citenamefont {Broggio}\ \emph {et~al.}(2017)\citenamefont
  {Broggio}, \citenamefont {Ferroglia}, \citenamefont {Pecjak},\ and\
  \citenamefont {Yang}}]{Broggio:2016lfj}%
  \BibitemOpen
  \bibfield  {author} {\bibinfo {author} {\bibfnamefont {A.}~\bibnamefont
  {Broggio}}, \bibinfo {author} {\bibfnamefont {A.}~\bibnamefont {Ferroglia}},
  \bibinfo {author} {\bibfnamefont {B.~D.}\ \bibnamefont {Pecjak}}, \ and\
  \bibinfo {author} {\bibfnamefont {L.~L.}\ \bibnamefont {Yang}},\ }\href
  {\doibase 10.1007/JHEP02(2017)126} {\bibfield  {journal} {\bibinfo  {journal}
  {JHEP}\ }\textbf {\bibinfo {volume} {02}},\ \bibinfo {pages} {126} (\bibinfo
  {year} {2017})},\ \Eprint {http://arxiv.org/abs/1611.00049} {arXiv:1611.00049
  [hep-ph]} \BibitemShut {NoStop}%
\bibitem [{\citenamefont {Pecjak}\ \emph {et~al.}(2016)\citenamefont {Pecjak},
  \citenamefont {Scott}, \citenamefont {Wang},\ and\ \citenamefont
  {Yang}}]{Pecjak:2016nee}%
  \BibitemOpen
  \bibfield  {author} {\bibinfo {author} {\bibfnamefont {B.~D.}\ \bibnamefont
  {Pecjak}}, \bibinfo {author} {\bibfnamefont {D.~J.}\ \bibnamefont {Scott}},
  \bibinfo {author} {\bibfnamefont {X.}~\bibnamefont {Wang}}, \ and\ \bibinfo
  {author} {\bibfnamefont {L.~L.}\ \bibnamefont {Yang}},\ }\href {\doibase
  10.1103/PhysRevLett.116.202001} {\bibfield  {journal} {\bibinfo  {journal}
  {Phys. Rev. Lett.}\ }\textbf {\bibinfo {volume} {116}},\ \bibinfo {pages}
  {202001} (\bibinfo {year} {2016})},\ \Eprint
  {http://arxiv.org/abs/1601.07020} {arXiv:1601.07020 [hep-ph]} \BibitemShut
  {NoStop}%
\bibitem [{\citenamefont {Czakon}\ \emph {et~al.}(2018)\citenamefont {Czakon},
  \citenamefont {Ferroglia}, \citenamefont {Heymes}, \citenamefont {Mitov},
  \citenamefont {Pecjak}, \citenamefont {Scott}, \citenamefont {Wang},\ and\
  \citenamefont {Yang}}]{Czakon:2018nun}%
  \BibitemOpen
  \bibfield  {author} {\bibinfo {author} {\bibfnamefont {M.}~\bibnamefont
  {Czakon}}, \bibinfo {author} {\bibfnamefont {A.}~\bibnamefont {Ferroglia}},
  \bibinfo {author} {\bibfnamefont {D.}~\bibnamefont {Heymes}}, \bibinfo
  {author} {\bibfnamefont {A.}~\bibnamefont {Mitov}}, \bibinfo {author}
  {\bibfnamefont {B.~D.}\ \bibnamefont {Pecjak}}, \bibinfo {author}
  {\bibfnamefont {D.~J.}\ \bibnamefont {Scott}}, \bibinfo {author}
  {\bibfnamefont {X.}~\bibnamefont {Wang}}, \ and\ \bibinfo {author}
  {\bibfnamefont {L.~L.}\ \bibnamefont {Yang}},\ }\href {\doibase
  10.1007/JHEP05(2018)149} {\bibfield  {journal} {\bibinfo  {journal} {JHEP}\
  }\textbf {\bibinfo {volume} {05}},\ \bibinfo {pages} {149} (\bibinfo {year}
  {2018})},\ \Eprint {http://arxiv.org/abs/1803.07623} {arXiv:1803.07623
  [hep-ph]} \BibitemShut {NoStop}%
\bibitem [{\citenamefont {Ju}\ \emph {et~al.}(2020)\citenamefont {Ju},
  \citenamefont {Wang}, \citenamefont {Wang}, \citenamefont {Xu}, \citenamefont
  {Xu},\ and\ \citenamefont {Yang}}]{Ju:2020otc}%
  \BibitemOpen
  \bibfield  {author} {\bibinfo {author} {\bibfnamefont {W.-L.}\ \bibnamefont
  {Ju}}, \bibinfo {author} {\bibfnamefont {G.}~\bibnamefont {Wang}}, \bibinfo
  {author} {\bibfnamefont {X.}~\bibnamefont {Wang}}, \bibinfo {author}
  {\bibfnamefont {X.}~\bibnamefont {Xu}}, \bibinfo {author} {\bibfnamefont
  {Y.}~\bibnamefont {Xu}}, \ and\ \bibinfo {author} {\bibfnamefont {L.~L.}\
  \bibnamefont {Yang}},\ }\href {\doibase 10.1007/JHEP06(2020)158} {\bibfield
  {journal} {\bibinfo  {journal} {JHEP}\ }\textbf {\bibinfo {volume} {06}},\
  \bibinfo {pages} {158} (\bibinfo {year} {2020})},\ \Eprint
  {http://arxiv.org/abs/2004.03088} {arXiv:2004.03088 [hep-ph]} \BibitemShut
  {NoStop}%
\bibitem [{\citenamefont {Bauer}\ \emph
  {et~al.}(2002{\natexlab{a}})\citenamefont {Bauer}, \citenamefont {Pirjol},\
  and\ \citenamefont {Stewart}}]{Bauer:2001yt}%
  \BibitemOpen
  \bibfield  {author} {\bibinfo {author} {\bibfnamefont {C.~W.}\ \bibnamefont
  {Bauer}}, \bibinfo {author} {\bibfnamefont {D.}~\bibnamefont {Pirjol}}, \
  and\ \bibinfo {author} {\bibfnamefont {I.~W.}\ \bibnamefont {Stewart}},\
  }\href {\doibase 10.1103/PhysRevD.65.054022} {\bibfield  {journal} {\bibinfo
  {journal} {Phys. Rev. D}\ }\textbf {\bibinfo {volume} {65}},\ \bibinfo
  {pages} {054022} (\bibinfo {year} {2002}{\natexlab{a}})},\ \Eprint
  {http://arxiv.org/abs/hep-ph/0109045} {arXiv:hep-ph/0109045} \BibitemShut
  {NoStop}%
\bibitem [{\citenamefont {Bauer}\ \emph
  {et~al.}(2002{\natexlab{b}})\citenamefont {Bauer}, \citenamefont {Fleming},
  \citenamefont {Pirjol}, \citenamefont {Rothstein},\ and\ \citenamefont
  {Stewart}}]{Bauer:2002nz}%
  \BibitemOpen
  \bibfield  {author} {\bibinfo {author} {\bibfnamefont {C.~W.}\ \bibnamefont
  {Bauer}}, \bibinfo {author} {\bibfnamefont {S.}~\bibnamefont {Fleming}},
  \bibinfo {author} {\bibfnamefont {D.}~\bibnamefont {Pirjol}}, \bibinfo
  {author} {\bibfnamefont {I.~Z.}\ \bibnamefont {Rothstein}}, \ and\ \bibinfo
  {author} {\bibfnamefont {I.~W.}\ \bibnamefont {Stewart}},\ }\href {\doibase
  10.1103/PhysRevD.66.014017} {\bibfield  {journal} {\bibinfo  {journal} {Phys.
  Rev. D}\ }\textbf {\bibinfo {volume} {66}},\ \bibinfo {pages} {014017}
  (\bibinfo {year} {2002}{\natexlab{b}})},\ \Eprint
  {http://arxiv.org/abs/hep-ph/0202088} {arXiv:hep-ph/0202088} \BibitemShut
  {NoStop}%
\bibitem [{\citenamefont {Beneke}\ \emph {et~al.}(2002)\citenamefont {Beneke},
  \citenamefont {Chapovsky}, \citenamefont {Diehl},\ and\ \citenamefont
  {Feldmann}}]{Beneke:2002ph}%
  \BibitemOpen
  \bibfield  {author} {\bibinfo {author} {\bibfnamefont {M.}~\bibnamefont
  {Beneke}}, \bibinfo {author} {\bibfnamefont {A.~P.}\ \bibnamefont
  {Chapovsky}}, \bibinfo {author} {\bibfnamefont {M.}~\bibnamefont {Diehl}}, \
  and\ \bibinfo {author} {\bibfnamefont {T.}~\bibnamefont {Feldmann}},\ }\href
  {\doibase 10.1016/S0550-3213(02)00687-9} {\bibfield  {journal} {\bibinfo
  {journal} {Nucl. Phys. B}\ }\textbf {\bibinfo {volume} {643}},\ \bibinfo
  {pages} {431} (\bibinfo {year} {2002})},\ \Eprint
  {http://arxiv.org/abs/hep-ph/0206152} {arXiv:hep-ph/0206152} \BibitemShut
  {NoStop}%
\bibitem [{Note1()}]{Note1}%
  \BibitemOpen
  \bibinfo {note} {Off-shellness $p_i^2$ is applied for the $i$th massless
  parton to regularize IR divergences in low-energy matrix
  elements.}\BibitemShut {Stop}%
\bibitem [{\citenamefont {Becher}\ \emph {et~al.}(2004)\citenamefont {Becher},
  \citenamefont {Hill}, \citenamefont {Lange},\ and\ \citenamefont
  {Neubert}}]{Becher:2003kh}%
  \BibitemOpen
  \bibfield  {author} {\bibinfo {author} {\bibfnamefont {T.}~\bibnamefont
  {Becher}}, \bibinfo {author} {\bibfnamefont {R.~J.}\ \bibnamefont {Hill}},
  \bibinfo {author} {\bibfnamefont {B.~O.}\ \bibnamefont {Lange}}, \ and\
  \bibinfo {author} {\bibfnamefont {M.}~\bibnamefont {Neubert}},\ }\href
  {\doibase 10.1103/PhysRevD.69.034013} {\bibfield  {journal} {\bibinfo
  {journal} {Phys. Rev. D}\ }\textbf {\bibinfo {volume} {69}},\ \bibinfo
  {pages} {034013} (\bibinfo {year} {2004})},\ \Eprint
  {http://arxiv.org/abs/hep-ph/0309227} {arXiv:hep-ph/0309227} \BibitemShut
  {NoStop}%
\bibitem [{\citenamefont {Catani}\ and\ \citenamefont
  {Seymour}(1996)}]{Catani:1996jh}%
  \BibitemOpen
  \bibfield  {author} {\bibinfo {author} {\bibfnamefont {S.}~\bibnamefont
  {Catani}}\ and\ \bibinfo {author} {\bibfnamefont {M.~H.}\ \bibnamefont
  {Seymour}},\ }\href {\doibase 10.1016/0370-2693(96)00425-X} {\bibfield
  {journal} {\bibinfo  {journal} {Phys. Lett. B}\ }\textbf {\bibinfo {volume}
  {378}},\ \bibinfo {pages} {287} (\bibinfo {year} {1996})},\ \Eprint
  {http://arxiv.org/abs/hep-ph/9602277} {arXiv:hep-ph/9602277} \BibitemShut
  {NoStop}%
\bibitem [{\citenamefont {Catani}\ and\ \citenamefont
  {Seymour}(1997)}]{Catani:1996vz}%
  \BibitemOpen
  \bibfield  {author} {\bibinfo {author} {\bibfnamefont {S.}~\bibnamefont
  {Catani}}\ and\ \bibinfo {author} {\bibfnamefont {M.~H.}\ \bibnamefont
  {Seymour}},\ }\href {\doibase 10.1016/S0550-3213(96)00589-5} {\bibfield
  {journal} {\bibinfo  {journal} {Nucl. Phys. B}\ }\textbf {\bibinfo {volume}
  {485}},\ \bibinfo {pages} {291} (\bibinfo {year} {1997})},\ \bibinfo {note}
  {[Erratum: Nucl.Phys.B 510, 503--504 (1998)]},\ \Eprint
  {http://arxiv.org/abs/hep-ph/9605323} {arXiv:hep-ph/9605323} \BibitemShut
  {NoStop}%
\bibitem [{\citenamefont {Gatheral}(1983)}]{Gatheral:1983cz}%
  \BibitemOpen
  \bibfield  {author} {\bibinfo {author} {\bibfnamefont {J.~G.~M.}\
  \bibnamefont {Gatheral}},\ }\href {\doibase 10.1016/0370-2693(83)90112-0}
  {\bibfield  {journal} {\bibinfo  {journal} {Phys. Lett. B}\ }\textbf
  {\bibinfo {volume} {133}},\ \bibinfo {pages} {90} (\bibinfo {year}
  {1983})}\BibitemShut {NoStop}%
\bibitem [{\citenamefont {Frenkel}\ and\ \citenamefont
  {Taylor}(1984)}]{Frenkel:1984pz}%
  \BibitemOpen
  \bibfield  {author} {\bibinfo {author} {\bibfnamefont {J.}~\bibnamefont
  {Frenkel}}\ and\ \bibinfo {author} {\bibfnamefont {J.~C.}\ \bibnamefont
  {Taylor}},\ }\href {\doibase 10.1016/0550-3213(84)90294-3} {\bibfield
  {journal} {\bibinfo  {journal} {Nucl. Phys. B}\ }\textbf {\bibinfo {volume}
  {246}},\ \bibinfo {pages} {231} (\bibinfo {year} {1984})}\BibitemShut
  {NoStop}%
\bibitem [{\citenamefont {Gardi}\ \emph {et~al.}(2010)\citenamefont {Gardi},
  \citenamefont {Laenen}, \citenamefont {Stavenga},\ and\ \citenamefont
  {White}}]{Gardi:2010rn}%
  \BibitemOpen
  \bibfield  {author} {\bibinfo {author} {\bibfnamefont {E.}~\bibnamefont
  {Gardi}}, \bibinfo {author} {\bibfnamefont {E.}~\bibnamefont {Laenen}},
  \bibinfo {author} {\bibfnamefont {G.}~\bibnamefont {Stavenga}}, \ and\
  \bibinfo {author} {\bibfnamefont {C.~D.}\ \bibnamefont {White}},\ }\href
  {\doibase 10.1007/JHEP11(2010)155} {\bibfield  {journal} {\bibinfo  {journal}
  {JHEP}\ }\textbf {\bibinfo {volume} {11}},\ \bibinfo {pages} {155} (\bibinfo
  {year} {2010})},\ \Eprint {http://arxiv.org/abs/1008.0098} {arXiv:1008.0098
  [hep-ph]} \BibitemShut {NoStop}%
\bibitem [{\citenamefont {Gardi}\ \emph {et~al.}(2013)\citenamefont {Gardi},
  \citenamefont {Smillie},\ and\ \citenamefont {White}}]{Gardi:2013ita}%
  \BibitemOpen
  \bibfield  {author} {\bibinfo {author} {\bibfnamefont {E.}~\bibnamefont
  {Gardi}}, \bibinfo {author} {\bibfnamefont {J.~M.}\ \bibnamefont {Smillie}},
  \ and\ \bibinfo {author} {\bibfnamefont {C.~D.}\ \bibnamefont {White}},\
  }\href {\doibase 10.1007/JHEP06(2013)088} {\bibfield  {journal} {\bibinfo
  {journal} {JHEP}\ }\textbf {\bibinfo {volume} {06}},\ \bibinfo {pages} {088}
  (\bibinfo {year} {2013})},\ \Eprint {http://arxiv.org/abs/1304.7040}
  {arXiv:1304.7040 [hep-ph]} \BibitemShut {NoStop}%
\bibitem [{Note2()}]{Note2}%
  \BibitemOpen
  \bibinfo {note} {Throughout this letter, the tripole correlation refers to
  full color connections of three partons, including both ${\protect \mathcal
  T}_{ijk}$ and ${\protect \mathcal T}_{iijk}$ up to three-loop order. This is
  different from the color tripole mentioned in \cite
  {Almelid:2015jia,Almelid:2017qju}}\BibitemShut {NoStop}%
\bibitem [{Note3()}]{Note3}%
  \BibitemOpen
  \bibinfo {note} {Simple Casimir scaling relation implies $\Gamma _{\protect
  \rm cusp}^{i}(\alpha _s) = C_{R_i}\gamma _{\protect \rm cusp}(\alpha _s)$,
  which is violated starting at four-loop order~\cite
  {Boels:2017skl,Boels:2017ftb}. Here $C_{R_i}={\protect \bm {T}}_i^2$ is the
  quadratic Casimir operator of parton $i$.}\BibitemShut {Stop}%
\bibitem [{\citenamefont {Moch}\ \emph {et~al.}(2004)\citenamefont {Moch},
  \citenamefont {Vermaseren},\ and\ \citenamefont {Vogt}}]{Moch:2004pa}%
  \BibitemOpen
  \bibfield  {author} {\bibinfo {author} {\bibfnamefont {S.}~\bibnamefont
  {Moch}}, \bibinfo {author} {\bibfnamefont {J.~A.~M.}\ \bibnamefont
  {Vermaseren}}, \ and\ \bibinfo {author} {\bibfnamefont {A.}~\bibnamefont
  {Vogt}},\ }\href {\doibase 10.1016/j.nuclphysb.2004.03.030} {\bibfield
  {journal} {\bibinfo  {journal} {Nucl. Phys. B}\ }\textbf {\bibinfo {volume}
  {688}},\ \bibinfo {pages} {101} (\bibinfo {year} {2004})},\ \Eprint
  {http://arxiv.org/abs/hep-ph/0403192} {arXiv:hep-ph/0403192} \BibitemShut
  {NoStop}%
\bibitem [{\citenamefont {Henn}\ \emph {et~al.}(2016)\citenamefont {Henn},
  \citenamefont {Smirnov}, \citenamefont {Smirnov},\ and\ \citenamefont
  {Steinhauser}}]{Henn:2016men}%
  \BibitemOpen
  \bibfield  {author} {\bibinfo {author} {\bibfnamefont {J.~M.}\ \bibnamefont
  {Henn}}, \bibinfo {author} {\bibfnamefont {A.~V.}\ \bibnamefont {Smirnov}},
  \bibinfo {author} {\bibfnamefont {V.~A.}\ \bibnamefont {Smirnov}}, \ and\
  \bibinfo {author} {\bibfnamefont {M.}~\bibnamefont {Steinhauser}},\ }\href
  {\doibase 10.1007/JHEP05(2016)066} {\bibfield  {journal} {\bibinfo  {journal}
  {JHEP}\ }\textbf {\bibinfo {volume} {05}},\ \bibinfo {pages} {066} (\bibinfo
  {year} {2016})},\ \Eprint {http://arxiv.org/abs/1604.03126} {arXiv:1604.03126
  [hep-ph]} \BibitemShut {NoStop}%
\bibitem [{\citenamefont {Davies}\ \emph {et~al.}(2017)\citenamefont {Davies},
  \citenamefont {Vogt}, \citenamefont {Ruijl}, \citenamefont {Ueda},\ and\
  \citenamefont {Vermaseren}}]{Davies:2016jie}%
  \BibitemOpen
  \bibfield  {author} {\bibinfo {author} {\bibfnamefont {J.}~\bibnamefont
  {Davies}}, \bibinfo {author} {\bibfnamefont {A.}~\bibnamefont {Vogt}},
  \bibinfo {author} {\bibfnamefont {B.}~\bibnamefont {Ruijl}}, \bibinfo
  {author} {\bibfnamefont {T.}~\bibnamefont {Ueda}}, \ and\ \bibinfo {author}
  {\bibfnamefont {J.~A.~M.}\ \bibnamefont {Vermaseren}},\ }\href {\doibase
  10.1016/j.nuclphysb.2016.12.012} {\bibfield  {journal} {\bibinfo  {journal}
  {Nucl. Phys. B}\ }\textbf {\bibinfo {volume} {915}},\ \bibinfo {pages} {335}
  (\bibinfo {year} {2017})},\ \Eprint {http://arxiv.org/abs/1610.07477}
  {arXiv:1610.07477 [hep-ph]} \BibitemShut {NoStop}%
\bibitem [{\citenamefont {Henn}\ \emph {et~al.}(2017)\citenamefont {Henn},
  \citenamefont {Smirnov}, \citenamefont {Smirnov}, \citenamefont
  {Steinhauser},\ and\ \citenamefont {Lee}}]{Henn:2016wlm}%
  \BibitemOpen
  \bibfield  {author} {\bibinfo {author} {\bibfnamefont {J.}~\bibnamefont
  {Henn}}, \bibinfo {author} {\bibfnamefont {A.~V.}\ \bibnamefont {Smirnov}},
  \bibinfo {author} {\bibfnamefont {V.~A.}\ \bibnamefont {Smirnov}}, \bibinfo
  {author} {\bibfnamefont {M.}~\bibnamefont {Steinhauser}}, \ and\ \bibinfo
  {author} {\bibfnamefont {R.~N.}\ \bibnamefont {Lee}},\ }\href {\doibase
  10.1007/JHEP03(2017)139} {\bibfield  {journal} {\bibinfo  {journal} {JHEP}\
  }\textbf {\bibinfo {volume} {03}},\ \bibinfo {pages} {139} (\bibinfo {year}
  {2017})},\ \Eprint {http://arxiv.org/abs/1612.04389} {arXiv:1612.04389
  [hep-ph]} \BibitemShut {NoStop}%
\bibitem [{\citenamefont {Lee}\ \emph {et~al.}(2017)\citenamefont {Lee},
  \citenamefont {Smirnov}, \citenamefont {Smirnov},\ and\ \citenamefont
  {Steinhauser}}]{Lee:2017mip}%
  \BibitemOpen
  \bibfield  {author} {\bibinfo {author} {\bibfnamefont {R.~N.}\ \bibnamefont
  {Lee}}, \bibinfo {author} {\bibfnamefont {A.~V.}\ \bibnamefont {Smirnov}},
  \bibinfo {author} {\bibfnamefont {V.~A.}\ \bibnamefont {Smirnov}}, \ and\
  \bibinfo {author} {\bibfnamefont {M.}~\bibnamefont {Steinhauser}},\ }\href
  {\doibase 10.1103/PhysRevD.96.014008} {\bibfield  {journal} {\bibinfo
  {journal} {Phys. Rev. D}\ }\textbf {\bibinfo {volume} {96}},\ \bibinfo
  {pages} {014008} (\bibinfo {year} {2017})},\ \Eprint
  {http://arxiv.org/abs/1705.06862} {arXiv:1705.06862 [hep-ph]} \BibitemShut
  {NoStop}%
\bibitem [{\citenamefont {Moch}\ \emph {et~al.}(2017)\citenamefont {Moch},
  \citenamefont {Ruijl}, \citenamefont {Ueda}, \citenamefont {Vermaseren},\
  and\ \citenamefont {Vogt}}]{Moch:2017uml}%
  \BibitemOpen
  \bibfield  {author} {\bibinfo {author} {\bibfnamefont {S.}~\bibnamefont
  {Moch}}, \bibinfo {author} {\bibfnamefont {B.}~\bibnamefont {Ruijl}},
  \bibinfo {author} {\bibfnamefont {T.}~\bibnamefont {Ueda}}, \bibinfo {author}
  {\bibfnamefont {J.~A.~M.}\ \bibnamefont {Vermaseren}}, \ and\ \bibinfo
  {author} {\bibfnamefont {A.}~\bibnamefont {Vogt}},\ }\href {\doibase
  10.1007/JHEP10(2017)041} {\bibfield  {journal} {\bibinfo  {journal} {JHEP}\
  }\textbf {\bibinfo {volume} {10}},\ \bibinfo {pages} {041} (\bibinfo {year}
  {2017})},\ \Eprint {http://arxiv.org/abs/1707.08315} {arXiv:1707.08315
  [hep-ph]} \BibitemShut {NoStop}%
\bibitem [{\citenamefont {Grozin}(2018)}]{Grozin:2018vdn}%
  \BibitemOpen
  \bibfield  {author} {\bibinfo {author} {\bibfnamefont {A.}~\bibnamefont
  {Grozin}},\ }\href {\doibase 10.1007/JHEP01(2019)134} {\bibfield  {journal}
  {\bibinfo  {journal} {JHEP}\ }\textbf {\bibinfo {volume} {06}},\ \bibinfo
  {pages} {073} (\bibinfo {year} {2018})},\ \bibinfo {note} {[Addendum: JHEP
  01, 134 (2019)]},\ \Eprint {http://arxiv.org/abs/1805.05050}
  {arXiv:1805.05050 [hep-ph]} \BibitemShut {NoStop}%
\bibitem [{\citenamefont {Moch}\ \emph {et~al.}(2018)\citenamefont {Moch},
  \citenamefont {Ruijl}, \citenamefont {Ueda}, \citenamefont {Vermaseren},\
  and\ \citenamefont {Vogt}}]{Moch:2018wjh}%
  \BibitemOpen
  \bibfield  {author} {\bibinfo {author} {\bibfnamefont {S.}~\bibnamefont
  {Moch}}, \bibinfo {author} {\bibfnamefont {B.}~\bibnamefont {Ruijl}},
  \bibinfo {author} {\bibfnamefont {T.}~\bibnamefont {Ueda}}, \bibinfo {author}
  {\bibfnamefont {J.~A.~M.}\ \bibnamefont {Vermaseren}}, \ and\ \bibinfo
  {author} {\bibfnamefont {A.}~\bibnamefont {Vogt}},\ }\href {\doibase
  10.1016/j.physletb.2018.06.017} {\bibfield  {journal} {\bibinfo  {journal}
  {Phys. Lett. B}\ }\textbf {\bibinfo {volume} {782}},\ \bibinfo {pages} {627}
  (\bibinfo {year} {2018})},\ \Eprint {http://arxiv.org/abs/1805.09638}
  {arXiv:1805.09638 [hep-ph]} \BibitemShut {NoStop}%
\bibitem [{\citenamefont {Lee}\ \emph {et~al.}(2019)\citenamefont {Lee},
  \citenamefont {Smirnov}, \citenamefont {Smirnov},\ and\ \citenamefont
  {Steinhauser}}]{Lee:2019zop}%
  \BibitemOpen
  \bibfield  {author} {\bibinfo {author} {\bibfnamefont {R.~N.}\ \bibnamefont
  {Lee}}, \bibinfo {author} {\bibfnamefont {A.~V.}\ \bibnamefont {Smirnov}},
  \bibinfo {author} {\bibfnamefont {V.~A.}\ \bibnamefont {Smirnov}}, \ and\
  \bibinfo {author} {\bibfnamefont {M.}~\bibnamefont {Steinhauser}},\ }\href
  {\doibase 10.1007/JHEP02(2019)172} {\bibfield  {journal} {\bibinfo  {journal}
  {JHEP}\ }\textbf {\bibinfo {volume} {02}},\ \bibinfo {pages} {172} (\bibinfo
  {year} {2019})},\ \Eprint {http://arxiv.org/abs/1901.02898} {arXiv:1901.02898
  [hep-ph]} \BibitemShut {NoStop}%
\bibitem [{\citenamefont {Henn}\ \emph {et~al.}(2019)\citenamefont {Henn},
  \citenamefont {Peraro}, \citenamefont {Stahlhofen},\ and\ \citenamefont
  {Wasser}}]{Henn:2019rmi}%
  \BibitemOpen
  \bibfield  {author} {\bibinfo {author} {\bibfnamefont {J.~M.}\ \bibnamefont
  {Henn}}, \bibinfo {author} {\bibfnamefont {T.}~\bibnamefont {Peraro}},
  \bibinfo {author} {\bibfnamefont {M.}~\bibnamefont {Stahlhofen}}, \ and\
  \bibinfo {author} {\bibfnamefont {P.}~\bibnamefont {Wasser}},\ }\href
  {\doibase 10.1103/PhysRevLett.122.201602} {\bibfield  {journal} {\bibinfo
  {journal} {Phys. Rev. Lett.}\ }\textbf {\bibinfo {volume} {122}},\ \bibinfo
  {pages} {201602} (\bibinfo {year} {2019})},\ \Eprint
  {http://arxiv.org/abs/1901.03693} {arXiv:1901.03693 [hep-ph]} \BibitemShut
  {NoStop}%
\bibitem [{\citenamefont {von Manteuffel}\ and\ \citenamefont
  {Schabinger}(2019)}]{vonManteuffel:2019wbj}%
  \BibitemOpen
  \bibfield  {author} {\bibinfo {author} {\bibfnamefont {A.}~\bibnamefont {von
  Manteuffel}}\ and\ \bibinfo {author} {\bibfnamefont {R.~M.}\ \bibnamefont
  {Schabinger}},\ }\href {\doibase 10.1103/PhysRevD.99.094014} {\bibfield
  {journal} {\bibinfo  {journal} {Phys. Rev. D}\ }\textbf {\bibinfo {volume}
  {99}},\ \bibinfo {pages} {094014} (\bibinfo {year} {2019})},\ \Eprint
  {http://arxiv.org/abs/1902.08208} {arXiv:1902.08208 [hep-ph]} \BibitemShut
  {NoStop}%
\bibitem [{\citenamefont {Henn}\ \emph
  {et~al.}(2020{\natexlab{a}})\citenamefont {Henn}, \citenamefont
  {Korchemsky},\ and\ \citenamefont {Mistlberger}}]{Henn:2019swt}%
  \BibitemOpen
  \bibfield  {author} {\bibinfo {author} {\bibfnamefont {J.~M.}\ \bibnamefont
  {Henn}}, \bibinfo {author} {\bibfnamefont {G.~P.}\ \bibnamefont
  {Korchemsky}}, \ and\ \bibinfo {author} {\bibfnamefont {B.}~\bibnamefont
  {Mistlberger}},\ }\href {\doibase 10.1007/JHEP04(2020)018} {\bibfield
  {journal} {\bibinfo  {journal} {JHEP}\ }\textbf {\bibinfo {volume} {04}},\
  \bibinfo {pages} {018} (\bibinfo {year} {2020}{\natexlab{a}})},\ \Eprint
  {http://arxiv.org/abs/1911.10174} {arXiv:1911.10174 [hep-th]} \BibitemShut
  {NoStop}%
\bibitem [{\citenamefont {von Manteuffel}\ \emph {et~al.}(2020)\citenamefont
  {von Manteuffel}, \citenamefont {Panzer},\ and\ \citenamefont
  {Schabinger}}]{vonManteuffel:2020vjv}%
  \BibitemOpen
  \bibfield  {author} {\bibinfo {author} {\bibfnamefont {A.}~\bibnamefont {von
  Manteuffel}}, \bibinfo {author} {\bibfnamefont {E.}~\bibnamefont {Panzer}}, \
  and\ \bibinfo {author} {\bibfnamefont {R.~M.}\ \bibnamefont {Schabinger}},\
  }\href {\doibase 10.1103/PhysRevLett.124.162001} {\bibfield  {journal}
  {\bibinfo  {journal} {Phys. Rev. Lett.}\ }\textbf {\bibinfo {volume} {124}},\
  \bibinfo {pages} {162001} (\bibinfo {year} {2020})},\ \Eprint
  {http://arxiv.org/abs/2002.04617} {arXiv:2002.04617 [hep-ph]} \BibitemShut
  {NoStop}%
\bibitem [{\citenamefont {Agarwal}\ \emph
  {et~al.}(2021{\natexlab{a}})\citenamefont {Agarwal}, \citenamefont {von
  Manteuffel}, \citenamefont {Panzer},\ and\ \citenamefont
  {Schabinger}}]{Agarwal:2021zft}%
  \BibitemOpen
  \bibfield  {author} {\bibinfo {author} {\bibfnamefont {B.}~\bibnamefont
  {Agarwal}}, \bibinfo {author} {\bibfnamefont {A.}~\bibnamefont {von
  Manteuffel}}, \bibinfo {author} {\bibfnamefont {E.}~\bibnamefont {Panzer}}, \
  and\ \bibinfo {author} {\bibfnamefont {R.~M.}\ \bibnamefont {Schabinger}},\
  }\href {\doibase 10.1016/j.physletb.2021.136503} {\bibfield  {journal}
  {\bibinfo  {journal} {Phys. Lett. B}\ }\textbf {\bibinfo {volume} {820}},\
  \bibinfo {pages} {136503} (\bibinfo {year} {2021}{\natexlab{a}})},\ \Eprint
  {http://arxiv.org/abs/2102.09725} {arXiv:2102.09725 [hep-ph]} \BibitemShut
  {NoStop}%
\bibitem [{\citenamefont {Kidonakis}(2009)}]{Kidonakis:2009ev}%
  \BibitemOpen
  \bibfield  {author} {\bibinfo {author} {\bibfnamefont {N.}~\bibnamefont
  {Kidonakis}},\ }\href {\doibase 10.1103/PhysRevLett.102.232003} {\bibfield
  {journal} {\bibinfo  {journal} {Phys. Rev. Lett.}\ }\textbf {\bibinfo
  {volume} {102}},\ \bibinfo {pages} {232003} (\bibinfo {year} {2009})},\
  \Eprint {http://arxiv.org/abs/0903.2561} {arXiv:0903.2561 [hep-ph]}
  \BibitemShut {NoStop}%
\bibitem [{\citenamefont {Grozin}\ \emph {et~al.}(2015)\citenamefont {Grozin},
  \citenamefont {Henn}, \citenamefont {Korchemsky},\ and\ \citenamefont
  {Marquard}}]{Grozin:2014hna}%
  \BibitemOpen
  \bibfield  {author} {\bibinfo {author} {\bibfnamefont {A.}~\bibnamefont
  {Grozin}}, \bibinfo {author} {\bibfnamefont {J.~M.}\ \bibnamefont {Henn}},
  \bibinfo {author} {\bibfnamefont {G.~P.}\ \bibnamefont {Korchemsky}}, \ and\
  \bibinfo {author} {\bibfnamefont {P.}~\bibnamefont {Marquard}},\ }\href
  {\doibase 10.1103/PhysRevLett.114.062006} {\bibfield  {journal} {\bibinfo
  {journal} {Phys. Rev. Lett.}\ }\textbf {\bibinfo {volume} {114}},\ \bibinfo
  {pages} {062006} (\bibinfo {year} {2015})},\ \Eprint
  {http://arxiv.org/abs/1409.0023} {arXiv:1409.0023 [hep-ph]} \BibitemShut
  {NoStop}%
\bibitem [{\citenamefont {Grozin}\ \emph {et~al.}(2016)\citenamefont {Grozin},
  \citenamefont {Henn}, \citenamefont {Korchemsky},\ and\ \citenamefont
  {Marquard}}]{Grozin:2015kna}%
  \BibitemOpen
  \bibfield  {author} {\bibinfo {author} {\bibfnamefont {A.}~\bibnamefont
  {Grozin}}, \bibinfo {author} {\bibfnamefont {J.~M.}\ \bibnamefont {Henn}},
  \bibinfo {author} {\bibfnamefont {G.~P.}\ \bibnamefont {Korchemsky}}, \ and\
  \bibinfo {author} {\bibfnamefont {P.}~\bibnamefont {Marquard}},\ }\href
  {\doibase 10.1007/JHEP01(2016)140} {\bibfield  {journal} {\bibinfo  {journal}
  {JHEP}\ }\textbf {\bibinfo {volume} {01}},\ \bibinfo {pages} {140} (\bibinfo
  {year} {2016})},\ \Eprint {http://arxiv.org/abs/1510.07803} {arXiv:1510.07803
  [hep-ph]} \BibitemShut {NoStop}%
\bibitem [{\citenamefont {Br\"user}\ \emph {et~al.}(2019)\citenamefont
  {Br\"user}, \citenamefont {Grozin}, \citenamefont {Henn},\ and\ \citenamefont
  {Stahlhofen}}]{Bruser:2019auj}%
  \BibitemOpen
  \bibfield  {author} {\bibinfo {author} {\bibfnamefont {R.}~\bibnamefont
  {Br\"user}}, \bibinfo {author} {\bibfnamefont {A.}~\bibnamefont {Grozin}},
  \bibinfo {author} {\bibfnamefont {J.~M.}\ \bibnamefont {Henn}}, \ and\
  \bibinfo {author} {\bibfnamefont {M.}~\bibnamefont {Stahlhofen}},\ }\href
  {\doibase 10.1007/JHEP05(2019)186} {\bibfield  {journal} {\bibinfo  {journal}
  {JHEP}\ }\textbf {\bibinfo {volume} {05}},\ \bibinfo {pages} {186} (\bibinfo
  {year} {2019})},\ \Eprint {http://arxiv.org/abs/1902.05076} {arXiv:1902.05076
  [hep-ph]} \BibitemShut {NoStop}%
\bibitem [{\citenamefont {Br\"user}\ \emph {et~al.}(2021)\citenamefont
  {Br\"user}, \citenamefont {Dlapa}, \citenamefont {Henn},\ and\ \citenamefont
  {Yan}}]{Bruser:2020bsh}%
  \BibitemOpen
  \bibfield  {author} {\bibinfo {author} {\bibfnamefont {R.}~\bibnamefont
  {Br\"user}}, \bibinfo {author} {\bibfnamefont {C.}~\bibnamefont {Dlapa}},
  \bibinfo {author} {\bibfnamefont {J.~M.}\ \bibnamefont {Henn}}, \ and\
  \bibinfo {author} {\bibfnamefont {K.}~\bibnamefont {Yan}},\ }\href {\doibase
  10.1103/PhysRevLett.126.021601} {\bibfield  {journal} {\bibinfo  {journal}
  {Phys. Rev. Lett.}\ }\textbf {\bibinfo {volume} {126}},\ \bibinfo {pages}
  {021601} (\bibinfo {year} {2021})},\ \Eprint
  {http://arxiv.org/abs/2007.04851} {arXiv:2007.04851 [hep-th]} \BibitemShut
  {NoStop}%
\bibitem [{\citenamefont {Moch}\ \emph
  {et~al.}(2005{\natexlab{a}})\citenamefont {Moch}, \citenamefont
  {Vermaseren},\ and\ \citenamefont {Vogt}}]{Moch:2005id}%
  \BibitemOpen
  \bibfield  {author} {\bibinfo {author} {\bibfnamefont {S.}~\bibnamefont
  {Moch}}, \bibinfo {author} {\bibfnamefont {J.~A.~M.}\ \bibnamefont
  {Vermaseren}}, \ and\ \bibinfo {author} {\bibfnamefont {A.}~\bibnamefont
  {Vogt}},\ }\href {\doibase 10.1088/1126-6708/2005/08/049} {\bibfield
  {journal} {\bibinfo  {journal} {JHEP}\ }\textbf {\bibinfo {volume} {08}},\
  \bibinfo {pages} {049} (\bibinfo {year} {2005}{\natexlab{a}})},\ \Eprint
  {http://arxiv.org/abs/hep-ph/0507039} {arXiv:hep-ph/0507039} \BibitemShut
  {NoStop}%
\bibitem [{\citenamefont {Moch}\ \emph
  {et~al.}(2005{\natexlab{b}})\citenamefont {Moch}, \citenamefont
  {Vermaseren},\ and\ \citenamefont {Vogt}}]{Moch:2005tm}%
  \BibitemOpen
  \bibfield  {author} {\bibinfo {author} {\bibfnamefont {S.}~\bibnamefont
  {Moch}}, \bibinfo {author} {\bibfnamefont {J.~A.~M.}\ \bibnamefont
  {Vermaseren}}, \ and\ \bibinfo {author} {\bibfnamefont {A.}~\bibnamefont
  {Vogt}},\ }\href {\doibase 10.1016/j.physletb.2005.08.067} {\bibfield
  {journal} {\bibinfo  {journal} {Phys. Lett. B}\ }\textbf {\bibinfo {volume}
  {625}},\ \bibinfo {pages} {245} (\bibinfo {year} {2005}{\natexlab{b}})},\
  \Eprint {http://arxiv.org/abs/hep-ph/0508055} {arXiv:hep-ph/0508055}
  \BibitemShut {NoStop}%
\bibitem [{\citenamefont {Baikov}\ \emph {et~al.}(2009)\citenamefont {Baikov},
  \citenamefont {Chetyrkin}, \citenamefont {Smirnov}, \citenamefont {Smirnov},\
  and\ \citenamefont {Steinhauser}}]{Baikov:2009bg}%
  \BibitemOpen
  \bibfield  {author} {\bibinfo {author} {\bibfnamefont {P.~A.}\ \bibnamefont
  {Baikov}}, \bibinfo {author} {\bibfnamefont {K.~G.}\ \bibnamefont
  {Chetyrkin}}, \bibinfo {author} {\bibfnamefont {A.~V.}\ \bibnamefont
  {Smirnov}}, \bibinfo {author} {\bibfnamefont {V.~A.}\ \bibnamefont
  {Smirnov}}, \ and\ \bibinfo {author} {\bibfnamefont {M.}~\bibnamefont
  {Steinhauser}},\ }\href {\doibase 10.1103/PhysRevLett.102.212002} {\bibfield
  {journal} {\bibinfo  {journal} {Phys. Rev. Lett.}\ }\textbf {\bibinfo
  {volume} {102}},\ \bibinfo {pages} {212002} (\bibinfo {year} {2009})},\
  \Eprint {http://arxiv.org/abs/0902.3519} {arXiv:0902.3519 [hep-ph]}
  \BibitemShut {NoStop}%
\bibitem [{\citenamefont {Korchemsky}\ and\ \citenamefont
  {Radyushkin}(1987)}]{Korchemsky:1987wg}%
  \BibitemOpen
  \bibfield  {author} {\bibinfo {author} {\bibfnamefont {G.~P.}\ \bibnamefont
  {Korchemsky}}\ and\ \bibinfo {author} {\bibfnamefont {A.~V.}\ \bibnamefont
  {Radyushkin}},\ }\href {\doibase 10.1016/0550-3213(87)90277-X} {\bibfield
  {journal} {\bibinfo  {journal} {Nucl. Phys. B}\ }\textbf {\bibinfo {volume}
  {283}},\ \bibinfo {pages} {342} (\bibinfo {year} {1987})}\BibitemShut
  {NoStop}%
\bibitem [{\citenamefont {Korchemsky}\ and\ \citenamefont
  {Radyushkin}(1992)}]{Korchemsky:1991zp}%
  \BibitemOpen
  \bibfield  {author} {\bibinfo {author} {\bibfnamefont {G.~P.}\ \bibnamefont
  {Korchemsky}}\ and\ \bibinfo {author} {\bibfnamefont {A.~V.}\ \bibnamefont
  {Radyushkin}},\ }\href {\doibase 10.1016/0370-2693(92)90405-S} {\bibfield
  {journal} {\bibinfo  {journal} {Phys. Lett. B}\ }\textbf {\bibinfo {volume}
  {279}},\ \bibinfo {pages} {359} (\bibinfo {year} {1992})},\ \Eprint
  {http://arxiv.org/abs/hep-ph/9203222} {arXiv:hep-ph/9203222} \BibitemShut
  {NoStop}%
\bibitem [{\citenamefont {Br\"user}\ \emph {et~al.}(2020)\citenamefont
  {Br\"user}, \citenamefont {Liu},\ and\ \citenamefont
  {Stahlhofen}}]{Bruser:2019yjk}%
  \BibitemOpen
  \bibfield  {author} {\bibinfo {author} {\bibfnamefont {R.}~\bibnamefont
  {Br\"user}}, \bibinfo {author} {\bibfnamefont {Z.~L.}\ \bibnamefont {Liu}}, \
  and\ \bibinfo {author} {\bibfnamefont {M.}~\bibnamefont {Stahlhofen}},\
  }\href {\doibase 10.1007/JHEP03(2020)071} {\bibfield  {journal} {\bibinfo
  {journal} {JHEP}\ }\textbf {\bibinfo {volume} {03}},\ \bibinfo {pages} {071}
  (\bibinfo {year} {2020})},\ \Eprint {http://arxiv.org/abs/1911.04494}
  {arXiv:1911.04494 [hep-ph]} \BibitemShut {NoStop}%
\bibitem [{\citenamefont {Becher}\ and\ \citenamefont
  {Neubert}(2020)}]{Becher:2019avh}%
  \BibitemOpen
  \bibfield  {author} {\bibinfo {author} {\bibfnamefont {T.}~\bibnamefont
  {Becher}}\ and\ \bibinfo {author} {\bibfnamefont {M.}~\bibnamefont
  {Neubert}},\ }\href {\doibase 10.1007/JHEP01(2020)025} {\bibfield  {journal}
  {\bibinfo  {journal} {JHEP}\ }\textbf {\bibinfo {volume} {01}},\ \bibinfo
  {pages} {025} (\bibinfo {year} {2020})},\ \Eprint
  {http://arxiv.org/abs/1908.11379} {arXiv:1908.11379 [hep-ph]} \BibitemShut
  {NoStop}%
\bibitem [{\citenamefont {Mitov}\ and\ \citenamefont
  {Moch}(2007)}]{Mitov:2006xs}%
  \BibitemOpen
  \bibfield  {author} {\bibinfo {author} {\bibfnamefont {A.}~\bibnamefont
  {Mitov}}\ and\ \bibinfo {author} {\bibfnamefont {S.}~\bibnamefont {Moch}},\
  }\href {\doibase 10.1088/1126-6708/2007/05/001} {\bibfield  {journal}
  {\bibinfo  {journal} {JHEP}\ }\textbf {\bibinfo {volume} {05}},\ \bibinfo
  {pages} {001} (\bibinfo {year} {2007})},\ \Eprint
  {http://arxiv.org/abs/hep-ph/0612149} {arXiv:hep-ph/0612149} \BibitemShut
  {NoStop}%
\bibitem [{\citenamefont {Becher}\ and\ \citenamefont
  {Melnikov}(2007)}]{Becher:2007cu}%
  \BibitemOpen
  \bibfield  {author} {\bibinfo {author} {\bibfnamefont {T.}~\bibnamefont
  {Becher}}\ and\ \bibinfo {author} {\bibfnamefont {K.}~\bibnamefont
  {Melnikov}},\ }\href {\doibase 10.1088/1126-6708/2007/06/084} {\bibfield
  {journal} {\bibinfo  {journal} {JHEP}\ }\textbf {\bibinfo {volume} {06}},\
  \bibinfo {pages} {084} (\bibinfo {year} {2007})},\ \Eprint
  {http://arxiv.org/abs/0704.3582} {arXiv:0704.3582 [hep-ph]} \BibitemShut
  {NoStop}%
\bibitem [{\citenamefont {Gardi}(2014)}]{Gardi:2013saa}%
  \BibitemOpen
  \bibfield  {author} {\bibinfo {author} {\bibfnamefont {E.}~\bibnamefont
  {Gardi}},\ }\href {\doibase 10.1007/JHEP04(2014)044} {\bibfield  {journal}
  {\bibinfo  {journal} {JHEP}\ }\textbf {\bibinfo {volume} {04}},\ \bibinfo
  {pages} {044} (\bibinfo {year} {2014})},\ \Eprint
  {http://arxiv.org/abs/1310.5268} {arXiv:1310.5268 [hep-ph]} \BibitemShut
  {NoStop}%
\bibitem [{\citenamefont {Liu}\ and\ \citenamefont
  {Stahlhofen}(2021)}]{Liu:2020wmp}%
  \BibitemOpen
  \bibfield  {author} {\bibinfo {author} {\bibfnamefont {Z.~L.}\ \bibnamefont
  {Liu}}\ and\ \bibinfo {author} {\bibfnamefont {M.}~\bibnamefont
  {Stahlhofen}},\ }\href {\doibase 10.1007/JHEP02(2021)128} {\bibfield
  {journal} {\bibinfo  {journal} {JHEP}\ }\textbf {\bibinfo {volume} {02}},\
  \bibinfo {pages} {128} (\bibinfo {year} {2021})},\ \Eprint
  {http://arxiv.org/abs/2010.05861} {arXiv:2010.05861 [hep-ph]} \BibitemShut
  {NoStop}%
\bibitem [{Note4()}]{Note4}%
  \BibitemOpen
  \bibinfo {note} {The web mixing matrices are even available at four
  loops~\cite {Agarwal:2020nyc,Agarwal:2021him}}\BibitemShut {NoStop}%
\bibitem [{\citenamefont {Nogueira}(1993)}]{Nogueira:1991ex}%
  \BibitemOpen
  \bibfield  {author} {\bibinfo {author} {\bibfnamefont {P.}~\bibnamefont
  {Nogueira}},\ }\href {\doibase 10.1006/jcph.1993.1074} {\bibfield  {journal}
  {\bibinfo  {journal} {J. Comput. Phys.}\ }\textbf {\bibinfo {volume} {105}},\
  \bibinfo {pages} {279} (\bibinfo {year} {1993})}\BibitemShut {NoStop}%
\bibitem [{\citenamefont {Smirnov}\ and\ \citenamefont
  {Chuharev}(2020)}]{Smirnov:2019qkx}%
  \BibitemOpen
  \bibfield  {author} {\bibinfo {author} {\bibfnamefont {A.~V.}\ \bibnamefont
  {Smirnov}}\ and\ \bibinfo {author} {\bibfnamefont {F.~S.}\ \bibnamefont
  {Chuharev}},\ }\href {\doibase 10.1016/j.cpc.2019.106877} {\bibfield
  {journal} {\bibinfo  {journal} {Comput. Phys. Commun.}\ }\textbf {\bibinfo
  {volume} {247}},\ \bibinfo {pages} {106877} (\bibinfo {year} {2020})},\
  \Eprint {http://arxiv.org/abs/1901.07808} {arXiv:1901.07808 [hep-ph]}
  \BibitemShut {NoStop}%
\bibitem [{\citenamefont {Klappert}\ \emph {et~al.}(2021)\citenamefont
  {Klappert}, \citenamefont {Lange}, \citenamefont {Maierh\"ofer},\ and\
  \citenamefont {Usovitsch}}]{Klappert:2020nbg}%
  \BibitemOpen
  \bibfield  {author} {\bibinfo {author} {\bibfnamefont {J.}~\bibnamefont
  {Klappert}}, \bibinfo {author} {\bibfnamefont {F.}~\bibnamefont {Lange}},
  \bibinfo {author} {\bibfnamefont {P.}~\bibnamefont {Maierh\"ofer}}, \ and\
  \bibinfo {author} {\bibfnamefont {J.}~\bibnamefont {Usovitsch}},\ }\href
  {\doibase 10.1016/j.cpc.2021.108024} {\bibfield  {journal} {\bibinfo
  {journal} {Comput. Phys. Commun.}\ }\textbf {\bibinfo {volume} {266}},\
  \bibinfo {pages} {108024} (\bibinfo {year} {2021})},\ \Eprint
  {http://arxiv.org/abs/2008.06494} {arXiv:2008.06494 [hep-ph]} \BibitemShut
  {NoStop}%
\bibitem [{\citenamefont {Meyer}(2018)}]{Meyer:2017joq}%
  \BibitemOpen
  \bibfield  {author} {\bibinfo {author} {\bibfnamefont {C.}~\bibnamefont
  {Meyer}},\ }\href {\doibase 10.1016/j.cpc.2017.09.014} {\bibfield  {journal}
  {\bibinfo  {journal} {Comput. Phys. Commun.}\ }\textbf {\bibinfo {volume}
  {222}},\ \bibinfo {pages} {295} (\bibinfo {year} {2018})},\ \Eprint
  {http://arxiv.org/abs/1705.06252} {arXiv:1705.06252 [hep-ph]} \BibitemShut
  {NoStop}%
\bibitem [{\citenamefont {Henn}\ \emph
  {et~al.}(2020{\natexlab{b}})\citenamefont {Henn}, \citenamefont
  {Mistlberger}, \citenamefont {Smirnov},\ and\ \citenamefont
  {Wasser}}]{Henn:2020lye}%
  \BibitemOpen
  \bibfield  {author} {\bibinfo {author} {\bibfnamefont {J.}~\bibnamefont
  {Henn}}, \bibinfo {author} {\bibfnamefont {B.}~\bibnamefont {Mistlberger}},
  \bibinfo {author} {\bibfnamefont {V.~A.}\ \bibnamefont {Smirnov}}, \ and\
  \bibinfo {author} {\bibfnamefont {P.}~\bibnamefont {Wasser}},\ }\href
  {\doibase 10.1007/JHEP04(2020)167} {\bibfield  {journal} {\bibinfo  {journal}
  {JHEP}\ }\textbf {\bibinfo {volume} {04}},\ \bibinfo {pages} {167} (\bibinfo
  {year} {2020}{\natexlab{b}})},\ \Eprint {http://arxiv.org/abs/2002.09492}
  {arXiv:2002.09492 [hep-ph]} \BibitemShut {NoStop}%
\bibitem [{\citenamefont {Henn}(2013)}]{Henn:2013pwa}%
  \BibitemOpen
  \bibfield  {author} {\bibinfo {author} {\bibfnamefont {J.~M.}\ \bibnamefont
  {Henn}},\ }\href {\doibase 10.1103/PhysRevLett.110.251601} {\bibfield
  {journal} {\bibinfo  {journal} {Phys. Rev. Lett.}\ }\textbf {\bibinfo
  {volume} {110}},\ \bibinfo {pages} {251601} (\bibinfo {year} {2013})},\
  \Eprint {http://arxiv.org/abs/1304.1806} {arXiv:1304.1806 [hep-th]}
  \BibitemShut {NoStop}%
\bibitem [{\citenamefont {Aglietti}\ and\ \citenamefont
  {Bonciani}(2004)}]{Aglietti:2004tq}%
  \BibitemOpen
  \bibfield  {author} {\bibinfo {author} {\bibfnamefont {U.}~\bibnamefont
  {Aglietti}}\ and\ \bibinfo {author} {\bibfnamefont {R.}~\bibnamefont
  {Bonciani}},\ }\href {\doibase 10.1016/j.nuclphysb.2004.07.018} {\bibfield
  {journal} {\bibinfo  {journal} {Nucl. Phys. B}\ }\textbf {\bibinfo {volume}
  {698}},\ \bibinfo {pages} {277} (\bibinfo {year} {2004})},\ \Eprint
  {http://arxiv.org/abs/hep-ph/0401193} {arXiv:hep-ph/0401193} \BibitemShut
  {NoStop}%
\bibitem [{\citenamefont {Tarasov}(1996)}]{Tarasov:1996br}%
  \BibitemOpen
  \bibfield  {author} {\bibinfo {author} {\bibfnamefont {O.~V.}\ \bibnamefont
  {Tarasov}},\ }\href {\doibase 10.1103/PhysRevD.54.6479} {\bibfield  {journal}
  {\bibinfo  {journal} {Phys. Rev. D}\ }\textbf {\bibinfo {volume} {54}},\
  \bibinfo {pages} {6479} (\bibinfo {year} {1996})},\ \Eprint
  {http://arxiv.org/abs/hep-th/9606018} {arXiv:hep-th/9606018} \BibitemShut
  {NoStop}%
\bibitem [{\citenamefont {Lee}(2010)}]{Lee:2009dh}%
  \BibitemOpen
  \bibfield  {author} {\bibinfo {author} {\bibfnamefont {R.~N.}\ \bibnamefont
  {Lee}},\ }\href {\doibase 10.1016/j.nuclphysb.2009.12.025} {\bibfield
  {journal} {\bibinfo  {journal} {Nucl. Phys. B}\ }\textbf {\bibinfo {volume}
  {830}},\ \bibinfo {pages} {474} (\bibinfo {year} {2010})},\ \Eprint
  {http://arxiv.org/abs/0911.0252} {arXiv:0911.0252 [hep-ph]} \BibitemShut
  {NoStop}%
\bibitem [{\citenamefont {Lee}(2014)}]{Lee:2013mka}%
  \BibitemOpen
  \bibfield  {author} {\bibinfo {author} {\bibfnamefont {R.~N.}\ \bibnamefont
  {Lee}},\ }\href {\doibase 10.1088/1742-6596/523/1/012059} {\bibfield
  {journal} {\bibinfo  {journal} {J. Phys. Conf. Ser.}\ }\textbf {\bibinfo
  {volume} {523}},\ \bibinfo {pages} {012059} (\bibinfo {year} {2014})},\
  \Eprint {http://arxiv.org/abs/1310.1145} {arXiv:1310.1145 [hep-ph]}
  \BibitemShut {NoStop}%
\bibitem [{\citenamefont {Panzer}(2015)}]{Panzer:2014caa}%
  \BibitemOpen
  \bibfield  {author} {\bibinfo {author} {\bibfnamefont {E.}~\bibnamefont
  {Panzer}},\ }\href {\doibase 10.1016/j.cpc.2014.10.019} {\bibfield  {journal}
  {\bibinfo  {journal} {Comput. Phys. Commun.}\ }\textbf {\bibinfo {volume}
  {188}},\ \bibinfo {pages} {148} (\bibinfo {year} {2015})},\ \Eprint
  {http://arxiv.org/abs/1403.3385} {arXiv:1403.3385 [hep-th]} \BibitemShut
  {NoStop}%
\bibitem [{\citenamefont {Remiddi}\ and\ \citenamefont
  {Vermaseren}(2000)}]{Remiddi:1999ew}%
  \BibitemOpen
  \bibfield  {author} {\bibinfo {author} {\bibfnamefont {E.}~\bibnamefont
  {Remiddi}}\ and\ \bibinfo {author} {\bibfnamefont {J.~A.~M.}\ \bibnamefont
  {Vermaseren}},\ }\href {\doibase 10.1142/S0217751X00000367} {\bibfield
  {journal} {\bibinfo  {journal} {Int. J. Mod. Phys. A}\ }\textbf {\bibinfo
  {volume} {15}},\ \bibinfo {pages} {725} (\bibinfo {year} {2000})},\ \Eprint
  {http://arxiv.org/abs/hep-ph/9905237} {arXiv:hep-ph/9905237} \BibitemShut
  {NoStop}%
\bibitem [{\citenamefont {Maitre}(2006)}]{Maitre:2005uu}%
  \BibitemOpen
  \bibfield  {author} {\bibinfo {author} {\bibfnamefont {D.}~\bibnamefont
  {Maitre}},\ }\href {\doibase 10.1016/j.cpc.2005.10.008} {\bibfield  {journal}
  {\bibinfo  {journal} {Comput. Phys. Commun.}\ }\textbf {\bibinfo {volume}
  {174}},\ \bibinfo {pages} {222} (\bibinfo {year} {2006})},\ \Eprint
  {http://arxiv.org/abs/hep-ph/0507152} {arXiv:hep-ph/0507152} \BibitemShut
  {NoStop}%
\bibitem [{\citenamefont {Gardi}\ \emph {et~al.}(2021)\citenamefont {Gardi},
  \citenamefont {Harley}, \citenamefont {Lodin}, \citenamefont {Palusa},
  \citenamefont {Smillie}, \citenamefont {White},\ and\ \citenamefont
  {Yeomans}}]{Gardi:2021gzz}%
  \BibitemOpen
  \bibfield  {author} {\bibinfo {author} {\bibfnamefont {E.}~\bibnamefont
  {Gardi}}, \bibinfo {author} {\bibfnamefont {M.}~\bibnamefont {Harley}},
  \bibinfo {author} {\bibfnamefont {R.}~\bibnamefont {Lodin}}, \bibinfo
  {author} {\bibfnamefont {M.}~\bibnamefont {Palusa}}, \bibinfo {author}
  {\bibfnamefont {J.~M.}\ \bibnamefont {Smillie}}, \bibinfo {author}
  {\bibfnamefont {C.~D.}\ \bibnamefont {White}}, \ and\ \bibinfo {author}
  {\bibfnamefont {S.}~\bibnamefont {Yeomans}},\ }\href {\doibase
  10.1007/JHEP12(2021)018} {\bibfield  {journal} {\bibinfo  {journal} {JHEP}\
  }\textbf {\bibinfo {volume} {12}},\ \bibinfo {pages} {018} (\bibinfo {year}
  {2021})},\ \Eprint {http://arxiv.org/abs/2110.01685} {arXiv:2110.01685
  [hep-ph]} \BibitemShut {NoStop}%
\bibitem [{\citenamefont {Boels}\ \emph {et~al.}(2017)\citenamefont {Boels},
  \citenamefont {Huber},\ and\ \citenamefont {Yang}}]{Boels:2017skl}%
  \BibitemOpen
  \bibfield  {author} {\bibinfo {author} {\bibfnamefont {R.~H.}\ \bibnamefont
  {Boels}}, \bibinfo {author} {\bibfnamefont {T.}~\bibnamefont {Huber}}, \ and\
  \bibinfo {author} {\bibfnamefont {G.}~\bibnamefont {Yang}},\ }\href {\doibase
  10.1103/PhysRevLett.119.201601} {\bibfield  {journal} {\bibinfo  {journal}
  {Phys. Rev. Lett.}\ }\textbf {\bibinfo {volume} {119}},\ \bibinfo {pages}
  {201601} (\bibinfo {year} {2017})},\ \Eprint
  {http://arxiv.org/abs/1705.03444} {arXiv:1705.03444 [hep-th]} \BibitemShut
  {NoStop}%
\bibitem [{\citenamefont {Boels}\ \emph {et~al.}(2018)\citenamefont {Boels},
  \citenamefont {Huber},\ and\ \citenamefont {Yang}}]{Boels:2017ftb}%
  \BibitemOpen
  \bibfield  {author} {\bibinfo {author} {\bibfnamefont {R.~H.}\ \bibnamefont
  {Boels}}, \bibinfo {author} {\bibfnamefont {T.}~\bibnamefont {Huber}}, \ and\
  \bibinfo {author} {\bibfnamefont {G.}~\bibnamefont {Yang}},\ }\href {\doibase
  10.1007/JHEP01(2018)153} {\bibfield  {journal} {\bibinfo  {journal} {JHEP}\
  }\textbf {\bibinfo {volume} {01}},\ \bibinfo {pages} {153} (\bibinfo {year}
  {2018})},\ \Eprint {http://arxiv.org/abs/1711.08449} {arXiv:1711.08449
  [hep-th]} \BibitemShut {NoStop}%
\bibitem [{\citenamefont {Agarwal}\ \emph {et~al.}(2020)\citenamefont
  {Agarwal}, \citenamefont {Danish}, \citenamefont {Magnea}, \citenamefont
  {Pal},\ and\ \citenamefont {Tripathi}}]{Agarwal:2020nyc}%
  \BibitemOpen
  \bibfield  {author} {\bibinfo {author} {\bibfnamefont {N.}~\bibnamefont
  {Agarwal}}, \bibinfo {author} {\bibfnamefont {A.}~\bibnamefont {Danish}},
  \bibinfo {author} {\bibfnamefont {L.}~\bibnamefont {Magnea}}, \bibinfo
  {author} {\bibfnamefont {S.}~\bibnamefont {Pal}}, \ and\ \bibinfo {author}
  {\bibfnamefont {A.}~\bibnamefont {Tripathi}},\ }\href {\doibase
  10.1007/JHEP05(2020)128} {\bibfield  {journal} {\bibinfo  {journal} {JHEP}\
  }\textbf {\bibinfo {volume} {05}},\ \bibinfo {pages} {128} (\bibinfo {year}
  {2020})},\ \Eprint {http://arxiv.org/abs/2003.09714} {arXiv:2003.09714
  [hep-ph]} \BibitemShut {NoStop}%
\bibitem [{\citenamefont {Agarwal}\ \emph
  {et~al.}(2021{\natexlab{b}})\citenamefont {Agarwal}, \citenamefont {Magnea},
  \citenamefont {Pal},\ and\ \citenamefont {Tripathi}}]{Agarwal:2021him}%
  \BibitemOpen
  \bibfield  {author} {\bibinfo {author} {\bibfnamefont {N.}~\bibnamefont
  {Agarwal}}, \bibinfo {author} {\bibfnamefont {L.}~\bibnamefont {Magnea}},
  \bibinfo {author} {\bibfnamefont {S.}~\bibnamefont {Pal}}, \ and\ \bibinfo
  {author} {\bibfnamefont {A.}~\bibnamefont {Tripathi}},\ }\href {\doibase
  10.1007/JHEP03(2021)188} {\bibfield  {journal} {\bibinfo  {journal} {JHEP}\
  }\textbf {\bibinfo {volume} {03}},\ \bibinfo {pages} {188} (\bibinfo {year}
  {2021}{\natexlab{b}})},\ \Eprint {http://arxiv.org/abs/2102.03598}
  {arXiv:2102.03598 [hep-ph]} \BibitemShut {NoStop}%
\end{thebibliography}%


\onecolumngrid
\newpage
\appendix

\section*{Supplemental material}

\subsection{A. Constraints from soft-collinear factorization}\label{app:scfac}
According to the non-abelian exponentiation theorem, the soft anomalous dimension of multi-leg amplitudes with a massive parton could in principle receive contributions from the following structures starting at three loops:
\begin{gather}
\cT_{iijj}\beta_{ij}\,,  \quad \cT_{iijj}\,, \quad \cT_{jjII}\beta_{Ij}\,,  \quad \cT_{jjII}\,,  \quad
\cT_{iijk} \beta_{ij}\,,  \quad \cT_{iijk} \beta_{jk}\,,  \quad \cT_{iijk} \,,  \quad 
\cT_{ijII} \beta_{ij}\,,  \quad \cT_{ijII} \beta_{Ii}\,, 
 \nn\\
\cT_{iijI} \beta_{ij} \,, \quad  \cT_{iijI}  \beta_{Ii}\,,  \quad \cT_{iijI}  \beta_{Ij}\,, \quad 
\cT_{ijkl} \beta_{ij}\,, \quad \cT_{ijkI}\beta_{ij} \,, \quad  \cT_{ijkI}\beta_{Ij} \,, 
\\
\cT_{ijII} \bar F_{\rm h2}^{[A]}(r_{ijI})\,,  \quad  \cT_{iijI} \bar F_{\rm h2}^{[B]}(r_{ijI})\,,  \quad 
\cT_{ijkI} \bar F_{\rm h3}(r_{ijI}, r_{ikI}, r_{jkI})\,,\quad
\cT_{ijkl}\, \bar F_{4}(\beta_{ijkl}, \beta_{ijkl}-2\beta_{ilkj})\,,
\nn
\end{gather}
where symmetry properties of $\cT_{ijkl}$ have been taken into consideration. Because of soft-collinear factorization, only terms depending on conformal cross ratios or linearly dependening on cusp angles need to be considered. The soft anomalous dimension should take the form of a sum of above structures over all unordered tuples of distinct parton indices. By performing the sums over certain parton indices and applying color conservation together with~(\ref{eq:cTijIIrel}), we have additional relations as follows
\begin{align}\label{eq:cTrelbysum}
\sum_{(i,j,k)}\cT_{iijk}\beta_{ij}=&-\frac{1}{2}\sum_{(i,j)}\left(\frac{C_A^2}{8}\bmT_i \cdot \bmT_j +\cT_{iijj}\right)\beta_{ij}
+\frac{1}{2}\sum_{(i,j,k)}\cT_{iijk}\beta_{jk}-\frac{1}{2}\sum_{I}\sum_{(i,j)}\cT_{ijII}\beta_{ij}
+\frac{1}{8}\sum_{(i,j,k,l)}\cT_{ijkl}\beta_{ijkl} \nn
\\&
-\frac{1}{2}\sum_{(I,J)}\sum_{(i,j)}\cT_{ijIJ}\beta_{ij}\,, \\
\sum_{I}\sum_{(i,j)}\cT_{ijII}\beta_{Ii} =& - \sum_{I,i}\left(\frac{C_A^2}{8}\bmT_i \cdot \bmT_I +\cT_{iiII}\right)\beta_{Ii}-
\sum_{(I,J)}\sum_{i}\cT_{iJII}\beta_{Ii}\,, \label{eq:cTijIJbijrel}\\
\sum_{I}\sum_{(i,j)}\cT_{iiIj}\beta_{Ii} =& - \sum_{I,i}\left(\frac{C_A^2}{8}\bmT_i \cdot \bmT_I +\cT_{iiII}\right)\beta_{Ii}-
\sum_{(I,J)}\sum_{i}\cT_{iiIJ}\beta_{Ii}\,, \\
\sum_{I}\sum_{(i,j,k)}\cT_{ijkI}\beta_{ij}=&-\frac{1}{2}\sum_{(i,j)}\left(\frac{C_A^2}{8}\bmT_i \cdot \bmT_j +\cT_{iijj}\right)\beta_{ij}
-\frac{1}{2}\sum_{(i,j,k)}\cT_{iijk}\beta_{jk}-\frac{1}{2}\sum_I\sum_{(i,j)}\cT_{ijII}\beta_{ij}
-\frac{1}{8}\sum_{(i,j,k,l)}\cT_{ijkl}\beta_{ijkl} \nn
\\&
-\frac{1}{2}\sum_{(I,J)}\sum_{(i,j)}\cT_{ijIJ}\beta_{ij}\,, \label{eq:cTijklbijrel}\\
\sum_I\sum_{(i,j,k)}\cT_{ijkI}\beta_{Ij}=&\sum_{I,i}\left(\frac{C_A^2}{8}\bmT_i \cdot \bmT_I +\cT_{iiII}\right)\beta_{Ii}+
\sum_{I}\sum_{(i,j)}\cT_{iijI}\beta_{Ij} + \sum_{(I,J)}\sum_{i}\cT_{iiIJ}\beta_{Ii} - \sum_{(I,J)}\sum_{(i,j)}\cT_{ijIJ}\beta_{Ii}\,. \label{eq:cTijkIbIjrel}
\end{align}
The above relations together with~(\ref{eq:cTijIIrel}) and symmetry properties of $\cT_{ijkl}$ can reduce the general form of three-loop soft anomalous dimensions to
\begin{equation}
\begin{aligned}
\bar {\bm \Gamma}_{s}^{(3)} =&
\sum_{(i,j)}\frac{\bmT_i\cdot \bmT_j}{2} (\beta_{ij}\fb_1 +\hb_1)
+\sum_{I,j} \bmT_I\cdot \bmT_j\left[\left(\fb_1+\frac{C_A^2}{8}\fb_2 \right)\beta_{Ij}+\hb_2\right]
+\sum_i{\bar c}_i+\sum_I{\bar c}_I
\\&
+\sum_{(i,j)}\cT_{iijj}\left(\beta_{ij}\ \fb_3  + \hb_3 \right)
+\sum_{I,j}\cT_{IIjj}\left(\beta_{Ij}\ \fb_4  + \hb_4 \right)
+\sum_{(i,j,k)} \cT_{iijk}\left(\beta_{jk}\ \fb_5  + \hb_5 \right)
+\sum_I\sum_{(i,j)}\cT_{iijI}\ \beta_{Ij} \ \fb_6 
\\&
+\sum_I\sum_{(i,j)} \cT_{ijII}\left[\beta_{ij} \ \fb_7  + \bar F_{\rm h2}(r_{ijI})\right]
+\sum_{(i,j,k,l)} \cT_{ijkl} \left[\beta_{ijkl}\ \fb_8  + \bar F_{\rm 4}(\beta_{ijkI}, \beta_{ijkl}-2\beta_{ilkj})\right]
\\&
+\sum_I\sum_{(i,j,k)} \cT_{ijkI} \bar F_{\rm h3}(r_{ijI},r_{ikI},r_{jkI})
+\cdots\,.
\end{aligned}
\end{equation}
Again, here and below the ellipses denote the non-dipole contributions involving two or more massive partons. Taking derivative with respect to collinear logarithm $L_i$ and applying color conservation, we have
\begin{align}\label{eq:dgs3dli}
 \frac{\partial\bar{\bm\Gamma}_s^{(3)}}{\partial L_i}= & C_{R_i}\left( \frac{C_A^2 }{4}\fb_5 - \fb_1\right)
 +\frac{C_A^2}{8}\sum_{I}\bmT_I\cdot \bmT_i (\fb_2+2\fb_5-2\fb_7) + 2\sum_{j\neq i}\cT_{iijj}(\fb_3-\fb_5)
 + \sum_I \cT_{iiII}(\fb_4-2\fb_7)\nn
 \\&
+ \sum_I\sum_{j\neq i}\cT_{jjiI}(\fb_6-2\fb_5) +\cdots \,,
\end{align}
where $C_{R_i} = \bmT_i^2$ is the quadratic Casimir operator of the corresponding color representation.
Constraints from soft-collinear factorization require that~(\ref{eq:dgs3dli}) depends only on the representation of parton $i$, so we have 
\begin{gather}
\fb_4=2\fb_7=\fb_2+2\fb_3\,,\qquad \fb_5=\fb_3\,,\qquad \fb_6=2\fb_3\,.
\end{gather}
Then the general form of soft anomalous dimensions can be rewritten as
\begin{align}\label{eq:sgenform}
\bar {\bm \Gamma}_{s}^{(3)} =& \left( \sum_{(i,j)}\frac{\bmT_i\cdot \bmT_j}{2} \beta_{ij} 
+ \sum_{I,j} \bmT_I\cdot \bmT_j \,\beta_{Ij}\right)\left(\fb_1-\frac{C_A^2}{4}\fb_3\right) 
+ \sum_{i}\left[C_{R_i}\left(\frac{C_A^2 }{8} \hb_3 - \hb_1\right) + {\bar c}_i \right]
+ \sum_{I} {\bar c}_I
\nn\\&
+  \sum_{(i,j,k)}\cT_{iijk} (\hb_5-\hb_3) 
+\sum_{(i,j,k,l)}\cT_{ijkl}\left[\left(-\frac{\fb_3}{4}+\fb_8\right)\beta_{ijkl} + \bar F_4(\beta_{ijkl},\beta_{ijkl}-2\beta_{ilkj})\right]
\nn\\&
+ \sum_I\sum_{(i,j)}\cT_{ijII}\left[\bar F_{\rm h2} (r_{ijI}) + \frac{\fb_2}{2} \ln r_{ijI}  - (\hb_3+\hb_4) \right] 
+\sum_I\sum_{(i,j,k)}\cT_{ijkI}\left[-2\fb_3\ln r_{ijI} + \bar F_{\rm h3}(r_{ijI},r_{ikI},r_{jkI})\right]
\nn\\&
+\sum_{I,j} \bmT_I \cdot \bmT_j  \left[\hb_2-\hb_1+ \frac{C_A^2}{8}(\hb_3-\hb_4) \right]
+ \cdots\,,
\end{align}
where we have used the following relations derived from~(\ref{eq:cTijIJbijrel}), (\ref{eq:cTijklbijrel}) and (\ref{eq:cTijkIbIjrel})
\begin{align}
\sum_{I,j} \cT_{jjII}\beta_{Ij} + \frac{1}{2} \sum_I\sum_{(i,j)} \cT_{ijII} \beta_{ij}
 = - \frac{C_A^2}{8}\sum_{I,j} \bmT_I\cdot\bmT_j\beta_{Ij} + \frac{1}{2}\sum_I\sum_{(i,j)}\cT_{ijII}\ln r_{ijI} +\cdots \,,
\end{align}
and
\begin{align}
&\sum_{(i,j)}\cT_{iijj}\beta_{ij} + \sum_{(i,j,k)}\cT_{iijk}\beta_{jk} 
+2 \sum_{I,j}\cT_{IIjj}\beta_{Ij}  + \sum_I\sum_{(i,j)}\cT_{ijII}\beta_{ij}  
+2 \sum_I\sum_{(i,j)}\cT_{iijI}\beta_{Ij}  \nn\\
=& 
-\frac{C_A^2}{4}\left[\sum_{(i,j)}\frac{\bmT_i\cdot\bmT_j}{2}\beta_{ij}
+\sum_{I,j}\bmT_I\cdot\bmT_j \beta_{Ij}\right] 
- 2\sum_I\sum_{(i,j,k)}\cT_{ijkI}\ln r_{ijI}
 - \frac{1}{4}\sum_{(i,j,k,l)}\cT_{ijkl}\beta_{ijkl}
 +\cdots \,.
\end{align}
Equation~(\ref{eq:sgenform}) is consistent with the final structure of anomalous dimensions shown in~(\ref{eq:gammafinalform}), if the last term in~(\ref{eq:sgenform}) is not taken into account. Actually, it indeed vanishes due to the constraints from small-mass limits shown in next section. 

\subsection{B. Constraints from small-mass and two-particle collinear limits}\label{app:m0andcollim}
First, we discuss on-shell amplitudes in the limit of small parton masses, which means that the masses of the external partons are much smaller the characteristic hard scales $s_{ij}$, $s_{Ij}$ and $s_{IJ}$. Refs.~\cite{Mitov:2006xs,Becher:2007cu} provide factorization theorems to describe the relation between on-shell masssive and massless scattering amplitudes in small-mass limits. The masses of external partons can be regarded as regulators to the collinear singularities for the corresponding massless amplitude, which can be encoded in universal jet functions belonging to each external parton to all orders in perturbative theory. This implies that in ${\bm\Gamma}(\{\underline{p}\},\{\underline{m}\to 0\},\mu) - {\bm\Gamma}(\{\underline{p}\},\{\underline 0\},\mu)$, there is no color exchange between different external partons, i.e.~\cite{Becher:2009kw}
\begin{equation}\label{eq:m0constr}
 {\bm\Gamma}(\{\underline{p}\},\{\underline{m}\to 0\},\mu) - {\bm\Gamma}(\{\underline{p}\},\{\underline 0\},\mu)
 =\sum_I\left[C_{R_I}\gcusp(\alpha_s) \ln\frac{\mu}{m_I}+ \gamma^Q -\gamma^q\right]\,,
\end{equation}
where ${\bm\Gamma}(\{\underline{p}\},\{\underline 0\},\mu)$ is the anomalous dimension for massless multi-leg amplitudes, which is known at three-loop order~\cite{Almelid:2015jia}. Here we use the fact that only quarks can be massive in QCD. Now we can understand that the last term in~(\ref{eq:sgenform}) must be excluded in the final structure of anomalous dimension in~(\ref{eq:gammafinalform}).  In the small-mass limit, the cusp angle of two massive partons can be written as
\begin{equation}
\lim_{m\to 0}\beta_{IJ}=
\lim_{m\to 0}{\rm cosh}^{-1}\left(\frac{-s_{IJ}}{2m_I m_J}\right)
\simeq \ln\frac{\mu}{m_I} + \ln\frac{\mu}{m_J} -\ln\frac{\mu^2}{-s_{IJ}}\,.
\end{equation} 
With the fact $\gcusp (\beta,\alpha_s)\simeq \gcusp(\alpha_s)\beta$ when $\beta\to \infty$, the difference between massive and the corresponding massless amplitudes in small-mass limits is given by 
\begin{align}
{\bm\Gamma}(\{\underline{p}\},\{\underline{m}\to 0\},\mu) -& {\bm\Gamma}(\{\underline{p}\},\{\underline 0\},\mu) 
= \sum_{I,i}\bmT_I\cdot \bmT_i\ln\frac{m_I}{\mu}+ \sum_{(I,J)}\bmT_I\cdot \bmT_J\ln\frac{m_I}{\mu}
+\sum_I \left(\gamma^Q - \gamma^q\right) 
\nn\\
&+ \sum_{I}\sum_{(i,j)}\Big[ \cT_{ijII}F_{\rm h2}(0,\alpha_s) -\left(\cT_{ijII} + \cT_{iiIj} + \cT_{jjiI}\right)f(\alpha_s)\Big]
\\
&+\sum_{I}\sum_{(i,j,k)}\cT_{ijkI}\left[\lim_{v_I^2\to 0}F_{\rm h3}(r_{ijI},r_{ikI},r_{jkI},\alpha_s)
 -4 F_{4}(\beta_{ijkI},\beta_{ijkI}-2\beta_{kjiI},\alpha_s)\right] \,.\nn
\end{align}
Using color conservation and the relation in~(\ref{eq:cTijIIrel}), the constraint in~(\ref{eq:m0constr}) indicates that 
\begin{equation}\label{eq:consfromm0}
F_{\rm h2}(0,\alpha_s)=3f(\alpha_s) \qquad \mbox{and} \qquad
\lim_{v_I^2\to 0}F_{\rm h3}(r_{ijI},r_{ikI},r_{jkI},\alpha_s) = 2 f(\alpha_s) + 4 F_{4}(\beta_{ijkI},\beta_{ijkI}-2\beta_{kjiI},\alpha_s)\,.
\end{equation}

Two-particle collinear limits also strongly constrain the structure of anomalous dimensions in~(\ref{eq:gammafinalform}). For a scattering amplitude where massless particles $1$ and $2$ (both assumed to be outgoing) become collinear, the relevant conformal cross ratios behave as 
\begin{gather}
\omega_{ij} \equiv \beta_{12ij}\to -\infty\,,\qquad
\beta_{1ij2}\to 0\,,\qquad 
r_{12I} = 0\,,\qquad 
\beta_{1ijk}=\beta_{2ijk} \,.
\end{gather}
Using the anti-symmetric properties $F_4(x,y,\alpha_s)=-F_4(-x,y,\alpha_s)$ and $F_{\rm h3}(x,y,z,\alpha_s) = -F_{\rm h3}(y,x,z,\alpha_s) $, the anomalous dimension of $a\to 1+2$ splitting amplitude can be simplified as
\begin{equation}\label{eq:gspdev}
\begin{aligned}
{\bf \Gamma}_{\rm Sp}(\{p_1,p_2\},\mu) =&
{\bf \Gamma}\left(\{p_1,p_2,\dots,p_n\},\{\underline{m}\},\mu\right)
-{\bf \Gamma}\left(\{p_a,\dots,p_n\},\{\underline{m}\},\mu\right)\\
=&\gcusp (\alpha_s)\left[{\bm T_1}\cdot {\bm T_2}\left(\ln\frac{\mu^2}{-s_{12}}+ \ln[z(1-z)]\right)
+C_{R_1}\ln z +C_{R_2}\ln (1-z) \right]\\
&+ \left[\gamma^1(\alpha_s)+\gamma^2(\alpha_s)-\gamma^a(\alpha_s)\right]{\bf 1}
+\Big[f(\alpha_s)+ 4 F_{4}(\omega_{ij},\omega_{ij},\alpha_s)\Big]
\left(-\frac{C_A^2}{4}{\bm T_1}\cdot {\bm T_2}-2\cT_{1122}\right)\\
&+4\sum_{i\neq 1,2}\cT_{12ii}\Big[f(\alpha_s) -  2 F_{4}(\omega_{ij},\omega_{ij},\alpha_s)\Big]
+2\sum_{I}\cT_{12II}\Big[F_{{\rm h}2}(0,\alpha_s)-f(\alpha_s) - 4 F_{4}(\omega_{ij},\omega_{ij},\alpha_s)\Big]\\
&+2\sum_{I}\sum_{i\neq 1,2}\left(\cT_{12iI}+\cT_{21iI}\right)\Big[F_{{\rm h}3}(0,r_{1iI},r_{1iI},\alpha_s) - 4 F_{4}(\omega_{ij},\omega_{ij},\alpha_s)\Big]
+ \cdots \,,
\end{aligned}
\end{equation}
where the following identity has been employed
\begin{equation}
\sum_{(i,j)}^{i,j\neq 1,2}\cT_{12ij} = -\frac{C_A^2}{8} \bmT_1 \cdot \bmT_2
-\cT_{1122}-\sum_{i\neq 1,2}\cT_{12ii} - \sum_I \cT_{12II} 
-\sum_{I,i}^{i\neq 1,2}\left(\cT_{12iI}+\cT_{12Ii}\right)-\sum_{(I,J)}\cT_{12IJ}\,.
\end{equation}
Collinear factorization requires the anomalous dimension of the splitting amplitude to be independent of color generators for particles other than 1 and 2, so the last three terms in~(\ref{eq:gspdev}) must vanish. Then we have
\begin{equation}\label{eq:consfromsplit}
\lim_{\omega\to -\infty}F_4(\omega,\omega,\alpha_s)=\frac{f(\alpha_s)}{2}\,,\qquad
F_{\rm h2}(0,\alpha_s)=3 f(\alpha_s)\,,\qquad
F_{\rm h3}(0,r,r,\alpha_s) = 2 f(\alpha_s)\,,
\end{equation}
where the second relation is consistent with the constraint from small-mass limits in~(\ref{eq:consfromm0}).

\end{document}